\documentclass[12pt,letterpaper]{article}
\usepackage[english]{babel}
\usepackage[margin=2.5cm]{geometry} 
\usepackage{amsmath, amssymb, mathtools, tensor}
\usepackage{mathrsfs} 
\usepackage{graphicx, float, subcaption, adjustbox}
\usepackage[rightcaption]{sidecap}
\usepackage{array}
\usepackage{csquotes}
\usepackage{xcolor}
\usepackage[colorlinks=true, allcolors=blue]{hyperref}
\usepackage{booktabs} 
\usepackage{appendix}
\usepackage[backend=biber, style=nature, sorting=none, natbib=true]{biblatex}
\usepackage{amssymb}   
\usepackage{pifont}    
\newcommand{\blackheart}{\ding{170}} 

\title{\textbf{Can Fractional Time Operators Reproduce Gravitational-Wave Memory? A No-Go Result}}

\author{ Sercan Kaya${}^{\maltese}$\thanks{\texttt{sercan.kaya@metu.edu.tr}}\,\,\,and \,\,
  Bayram Tekin$\textcolor{black}{\textsuperscript{\blackheart}}$\thanks{\texttt{bayram.tekin@bilkent.edu.tr}} \\[10pt]
  ${}^{\maltese}$Department of Physics, Middle East Technical University, 06800 Ankara, Türkiye \\[4pt]
  $\textcolor{black}{\textsuperscript{\blackheart}}$Department of Physics, Bilkent University, 06800 Ankara, Türkiye
}

\date{\today}

\begin{document}
\maketitle

\bigskip

\begin{abstract}
\noindent We initiate an investigation into whether fractional calculus, with its intrinsic long-tailed memory and nonlocal features, can provide a viable model for gravitational-wave memory effects. We consider two toy constructions: ($i$) a fractional modification of the linearized Einstein field equations using a sequential Caputo operator; and ($ii$) a fractionalized quadrupole formula in which the same operator acts on the source moment. Both constructions yield history-dependent responses with small memory-like offsets. However, in all cases we studied, the signal decays to zero at late times, failing to reproduce the permanent displacement predicted by General Relativity. We showed that, under asymptotic and spatial flatness of spacetime, the solutions of the proposed models decay to zero at late times when the time derivatives of the perturbed metric are temporally localized and bounded at each spatial point. Therefore, our results constitute a no-go demonstration: naive fractionalization is insufficient to model the permanent offset in the metric without explicitly building in flux-balance laws or asymptotic symmetry structure. We argue that any successful model must incorporate fractional kernels directly into the hereditary flux-balance integral of General Relativity while preserving gauge invariance and dimensional consistency. We also discuss possible connections to modified gravity and the absence of memory in spacetime with $D>4$ dimensions.
\end{abstract}

\setcounter{tocdepth}{2}
\tableofcontents

\section{Introduction}

The nonlinear gravitational wave (GW) memory effect is fundamentally hereditary \cite{Christodoulou1991,WisemanWill1991,BlanchetDamour1992,Thorne1992}: the permanent offset in the metric is obtained by integrating the flux of radiated energy to null infinity. This hereditary nature is reminiscent of the long-tailed memory kernels that naturally arise in fractional calculus \cite{Podlubny1999,Herrmann2018}. Unlike integer-order differential operators, fractional operators encode a continuum of scales, making them attractive candidates for modeling gravitational systems with nonlocal, history-dependent responses. The central question of this work is whether fractional derivatives can provide a more natural or flexible mathematical framework for capturing such hereditary effects than the classical integral formulations of General Relativity (GR).
In this work, we use the term “memory-like” to refer to permanent or slowly decaying offsets in field solutions arising from hereditary operators. We emphasize that this should be distinguished from nonlinear gravitational-wave memory in GR, which is a radiative effect defined at null infinity and governed by flux-balance laws and asymptotic symmetries.

Gravitation with fractional calculus has been studied in several works \cite{Calcagni2021,Palacios2023,Teodoro2023,Contreras2025}. For modifications of the Einstein field equations proposed in these papers, one improvement would be to use a dimensionally consistent form of the fractional models (see Sec. \ref{Modification of the Linearized Einstein Field Equation}). To our knowledge, no prior studies have modeled gravitational memory using fractional derivatives.

The linear memory effect \cite{Zeldovich1974} arises in linearized GR from the flux of unbound matter. The nonlinear memory effect \cite{Christodoulou1991,BlanchetDamour1992} is sourced by the GWs themselves: the energy carried by the radiation produces a permanent change in the gravitational field after the burst passes. Thorne \cite{Thorne1992} provided a useful physical interpretation of the nonlinear memory: the gravitons emitted by the source can be regarded as the effective unbound radiation that produces the memory.

The recent surge of activity on the theory side of the memory effect is mainly due to its connection to the infrared structure of GR. The soft graviton theorem \cite{Strominger2014} shows that adding a low-energy graviton to a scattering process is equivalent to a Ward identity associated with the asymptotic symmetries at null infinity. These symmetries form the Bondi-van der Burg-Metzner-Sachs (BMS) group. One class of these symmetries, the supertranslations, appears physically as the permanent displacement in position that detectors would measure after a GW passes \cite{Strominger2017}. In this way, memory effects, soft theorems, and asymptotic symmetries constitute the “infrared triangle”. In addition to the displacement memory, several other memory effects have been discovered, one of which is spin memory related to superrotations and subleading soft theorems \cite{Pasterski2016}. Memory has also been studied in linearized massive gravity and in higher curvature and higher-dimensional gravity theories. For example, the relation between the memory effect and graviton mass has been investigated in \cite{KilicarslanTekin2019}, which shows that observations of GW memory can strictly bound the graviton mass or potentially be incompatible with a massive graviton. In higher-dimensional spacetimes with an even number of dimensions, memory is non-existent \cite{Garfinkle2017}. From the computational point of view, an elegant derivation of memory was given in \cite{Garfinkle2022}: it was shown that the gravitational wave memory arises as a consequence of the fact that the Riemann tensor obeys a wave equation (sometimes called the Penrose equation) for all Riemannian spacetimes, including the Einstein metrics.

In the far zone, nonlinear memory can be expressed as \cite{WisemanWill1991,Thorne1992}
\begin{equation}\label{eq: the nonlinear memory}
    \Delta h_{ij}^{TT}
    = \frac{4}{r}\int_{-\infty}^{\infty} du\,
      \left[ \int \frac{dE^{\mathrm{gw}}}{du\, d\Omega'}\,
      \frac{n'_i n'_j}{1-\boldsymbol{n}'\!\cdot\!\boldsymbol{N}}\, d\Omega' \right]^{TT},
\end{equation}
where $u$ is the retarded time, $\Omega'$ is the solid angle, $\boldsymbol{n}'$ is the unit vector in the direction $\Omega'$, and $\boldsymbol{N}$ is the detector direction. [While this equation is well established, in Appendix A, we have provided a  heuristic description of it for completeness.]

From an observational perspective, the detection of GW memory is anticipated in the near future with third-generation ground-based detectors (such as the Einstein Telescope and Cosmic Explorer) and with the space-based mission LISA \cite{Boersma2020,Huebner2019,Grant2023}. Observing memory would provide unique confirmation of the nonlinear nature of Einstein's equations in the radiative regime, offering direct evidence for the flux-balance laws associated with asymptotic symmetries \cite{Christodoulou1991,Strominger2014,Strominger2017}. In addition, measuring memory could yield new constraints on the modifications of gravity in the strong-field radiative sector, beyond what is accessible from the oscillatory components of the waveform alone \cite{KilicarslanTekin2019,Garfinkle2017}. Thus, theoretical models that attempt to capture memory effects—whether in standard GR or in generalized frameworks such as fractional calculus—are of immediate relevance to the upcoming era of gravitational astronomy. Importantly, even a negative result is valuable: it delineates which kinds of nonlocal or fractional structures cannot reproduce nonlinear memory, thereby sharpening our understanding of which mathematical frameworks remain compatible with the fundamental predictions of GR.

The layout of the paper is as follows: In Sec.\,\ref{Definitions and Classification}, we briefly review the definition and a working classification of fractional operators. In Sec.\,\ref{Discussion and Results}, we investigate two fractional modifications to assess their ability to produce a permanent offset. In Sec.\,\ref{Modification of the Linearized Einstein Field Equation}, we analyze the fractional linearized Einstein field equations with a sequential Caputo time-fractional derivative, while Sec.\,\ref{Fractional Quadrupole Moment} applies the same operator to the quadrupole moment. In Sec.\,\ref{Late-Time Behavior of Fractional Wave Model and Gravitational Memory}, we analyze the late-time behavior of the fractional wave model under certain restrictions. Details of the numerical scheme are provided in Appendix B.

\section{Definitions and Classification}\label{Definitions and Classification}
The first step in modeling physical systems with fractional derivatives is to choose the definition of the derivative for the model. Many definitions for fractional derivatives exist in the literature; see \cite{Oliveira2014} for a review and a brief historical discussion. Before deciding on a definition that is well-suited for our model, we briefly explain a proposed criterion for 'what makes an operator a fractional derivative' and a classification of such operators.

In \cite{Ortigueira2015}, a criterion denoted by $C_2$ was proposed in \cite{Teodoro2019} to determine whether an operator is a fractional derivative. They refined the earlier proposal and introduced two criteria: the broad-sense criterion and the strict-sense criterion. In their proposal, fractional derivatives should satisfy linearity, an identity property, backward compatibility, the index law (semigroup property \cite{Herrmann2018}), and the generalized Leibniz rule. They have also shown that the Grünwald-Letnikov, Riemann-Liouville, and Caputo fractional derivatives mostly satisfy these properties in the broad sense under appropriate conditions.

\newcolumntype{L}[1]{>{\raggedright\arraybackslash}p{#1}}
\newcolumntype{C}[1]{>{\centering\arraybackslash}p{#1}}

\begin{table}[htbp]
  \centering
  \caption{Classification of Non-integer Order Derivative Operators}
  \label{tab:frac-ops}
  \setlength{\tabcolsep}{6pt}
  \renewcommand{\arraystretch}{1.15}
  \begin{tabular}{@{} C{0.10\linewidth} L{0.33\linewidth} L{0.50\linewidth} @{}}
    \hline\hline
    \textbf{Class} & \textbf{Type of Operator} & \textbf{Brief Description / Typical Examples} \\
    \hline
    F1 & Classical \emph{fractional} derivatives
       & Nonlocal, weakly singular kernels; history dependent.
         Examples: Riemann–Liouville, Caputo, Grünwald–Letnikov.\\
    \addlinespace[2pt]
    F2 & Modified / regularized derivatives
       & Systematic variants of F1 to adjust initial data, temper long tails,
         or improve well-posedness.
         Examples: Weyl, Regularized Liouville, Erdélyi–Kober.\\
    \addlinespace[2pt]
    F3 & Local operators
       & Operators with local (pointwise) definitions that mimic some
         fractional-like scalings but lack true hereditary memory.
         Examples: “conformable’’ derivative, Katugampola.\\
    \addlinespace[2pt]
    F4 & Non-singular kernel operators
       & Integral operators with nonsingular kernels aimed at alleviating
         initial-time singularities.
         Examples: Caputo–Fabrizio (exponential kernel)\\
    \hline\hline
  \end{tabular}

  \vspace{4pt}
  \begin{minipage}{0.95\linewidth}\footnotesize
  \emph{Note.} F1 and F2 are genuinely nonlocal (hereditary) in time; F3 is local and thus
  not equivalent to classical fractional calculus; F4 replaces weakly singular kernels by
  non-singular ones, which change short-time behavior and sometimes the physical
  interpretation of “memory’’.
  \end{minipage}
\end{table}

\noindent For $F4$ class operators with a non-singular kernel, linearity is the only property universally satisfied under criterion $C_2$, and the remaining properties do not exhibit a uniform pattern \cite{Teodoro2019}.

There are a few properties that, in some sense, characterize the fractional derivative operators in terms of the laws that the fractional derivatives violate. However, new definitions were proposed in \cite{Khalil2014,Katugampola2014}, which retain some of the classical properties of integer-order derivatives. Tarasov argued that these definitions are not fractional derivatives, as a finite set of integer-order derivatives can represent them \cite{Tarasov2013}. Moreover, well-defined fractional derivatives were shown to be nonlocal and do not obey the classical chain rule \cite{Tarasov2016,Tarasov2018}. Consistent with these views, we exclude local operators ($F3$ class) and operators with non-singular kernels ($F4$ class) to model the nonlinear memory effect.

Since our models involve the wave equation and the quadrupole moment and require physically meaningful initial conditions, we use the left-sided Caputo fractional derivative \cite{Caputo1967}:
\begin{equation}\label{eq: Caputo fractional derivative definition}
  {}_{t_a}^{C}D_{t}^{\alpha} f(t):= \frac{1}{\Gamma(m-\alpha)}\int_{t_a}^{t} (t-\xi)^{m-\alpha-1} f^{(m)}(\xi) d\xi,
\end{equation}
where $m-1<\alpha<m\in\mathbb{N}$ and $\Gamma(\cdot)$ are the gamma functions. The left-sided form is causal and appropriate for the classical initial conditions \cite{Herrmann2018}.
Throughout this work, we restrict to $0<\alpha<1$ (i.e., $m=1$) and set the lower terminal to $t_a=0^+$ where $t_a\leqslant 0$ causes singularity issues for equations \eqref{eq: Fractional 1+3 EFE w dimensional consistency} and \eqref{eq: 1+1 Fractional Wave Equation}. We use the shorthand notation
\begin{equation}\label{eq: Caputo fractional derivative notation}
  \partial_t^{\alpha} f(t)
  := \frac{1}{\Gamma(1-\alpha)}
     \int_{0^+}^{t} (t-\xi)^{-\alpha} f'(\xi) d\xi,
\end{equation}
for the Caputo time-fractional derivative used in our models, where $0<\alpha<1$.

\section{Main Results}\label{Discussion and Results}
In Sec.\,\ref{Definitions and Classification}, we restricted the candidate definitions of the fractional derivative to a smaller, well-motivated set. Nevertheless, many definitions remain applicable, and the literature does not offer a specific recipe for choosing the appropriate definition for the physical system. We adopt the left-sided Caputo derivative because it preserves causality and physically meaningful initial conditions.

The reader may find a review of the use of fractional derivatives in \cite{Herrmann2018,Tarasov2011book}. Such models are typically used when the classical local theories fail to capture the observed physical behavior. Nonlocality and long-time memory are fundamental features of fractional models.

Although nonlinear memory has not yet been observed, GR predicts both linear and nonlinear memory effects. Thus, any viable fractional model must reproduce the GR predictions, at least in an appropriate limit. Hence, we compare our results with GR based on the qualitative features of nonlinear memory. After establishing consistency with GR, fractional formulations may offer new insights into memory effects.

\subsection{Modification of the linearized Einstein field equations}
\label{Modification of the Linearized Einstein Field Equation}

We start with the linearized Einstein Field Equations (EFE) in the Lorenz gauge,
\begin{equation}\label{eq: Linearized EFE}
  \Box \,\bar{h}_{\mu\nu} = -\,\frac{16\pi G}{c^{4}}\,T_{\mu\nu},
  \qquad \Box := -\,\frac{1}{c^{2}}\partial_{t}^{2} + \Delta ,
\end{equation}
where $\bar{h}_{\mu\nu}$ is the trace-reversed field.
A naive way to “fractionalize” time is
\begin{equation}\label{eq: Fractional 1+3 EFE w/o dimensional consistency}
  \bigl(-\frac{1}{c^2}\partial_{t}^{\alpha} + \Delta\bigr)\bar{h}_{\mu\nu}
  = -\frac{16\pi G}{c^{4}}T_{\mu\nu},
  \qquad 1<\alpha<2 .
\end{equation}
Equation \eqref{eq: Fractional 1+3 EFE w/o dimensional consistency} is the inhomogeneous Caputo time fractional diffusion–wave equation in $1{+}3$ dimensions. The existence and uniqueness of fractional differential equations are discussed (see \cite{Podlubny1999}); the fundamental solutions for the homogeneous case in $1{+}1$ dimensions are discussed in \cite{Mainardi1996}. However, \eqref{eq: Fractional 1+3 EFE w/o dimensional consistency} is dimensionally inconsistent, the time operator scales as $[T^{-\alpha}]$ whereas $\Delta$ scales as $[L^{-2}]$. One can restore dimensional consistency by introducing a time scale $\tau$,
\begin{equation}
  -\frac{\tau^{\alpha-2}}{c^2}\partial_{t}^{\alpha}\bar{h}_{\mu\nu} + \Delta\bar{h}_{\mu\nu}=-\frac{16\pi G}{c^{4}}T_{\mu\nu},    
\end{equation}
or by using the Caputo–type Erdélyi–Kober fractional derivative \cite{Luchko2007,Samko1993}. Motivated by \cite{Kavvas2017,Kavvas2022}, we consider a dimensionally consistent sequential fractional operator:
\begin{equation}\label{eq: Fractional 1+3 EFE w dimensional consistency}
  \left(-\frac{\Gamma(2-\alpha)}{c\,t^{1-\alpha}}\partial_{t}^{\alpha}\Bigl[\frac{\Gamma(2-\alpha)}{c\,t^{1-\alpha}}\partial_{t}^{\alpha}\Bigr]+\Delta\right)\bar{h}_{\mu\nu}= -\frac{16\pi G}{c^{4}}T_{\mu\nu},
  \qquad 0<\alpha<1,
\end{equation}
This sequential Caputo operator is dimensionally consistent and, as $\alpha\to 1^-$, converges to $\partial_t^2$ for $t>0$. However, the form of the Caputo derivative in \eqref{eq: Fractional 1+3 EFE w dimensional consistency} is not invariant under time translations; hence, the solutions of \eqref{eq: Fractional 1+3 EFE w dimensional consistency} depend on the choice of time origin. To avoid the singularity at $t=0$, we take the lower terminal to be $t_a=0^+$ for the Caputo definition \eqref{eq: Caputo fractional derivative notation}.

To put the main question in context, after one splits the metric as $g_{\mu\nu}=\eta_{\mu\nu}+h_{\mu\nu}$ where $\eta_{\mu\nu}$ is the flat metric and $h_{\mu\nu}$ is the perturbed metric associated with the gravitational waves. At the next order in a perturbative expansion, the quadratic terms in $h_{\mu\nu}$ can be moved to the right-hand side and interpreted as the energy and momentum carried by the gravitational waves and, therefore, in the Lorenz gauge,

\begin{equation}
    \Box \bar{h}_{\mu\nu}
    =
    -\frac{16\pi G}{c^4}\bigl( T_{\mu\nu}+t_{\mu\nu}\bigr).
\end{equation}
Here, $t_{\mu\nu}=\frac{c^4}{32\pi G}{\left\langle \partial_\mu h_{\alpha\beta} \partial_\nu h^{\alpha \beta} \right\rangle}$ denotes the effective stress-energy of the gravitational waves. (See Appendix A for how memory is related to the effective stress-energy of GWs.)

The purpose of introducing the fractional model \eqref{eq: Fractional 1+3 EFE w dimensional consistency} is to investigate whether the nonlocality of the Caputo operator \eqref{eq: Caputo fractional derivative definition} can produce the nonlinear memory effects without explicitly constructing $t_{\mu\nu}$ and decomposing the metric as background and perturbed metric. To be more precise, we are not claiming this as the nonlinear memory of GR; rather, we are simply studying memory-like behavior that arises within this fractional model.

Even if the resulting equation is linear in the field, the response is history-dependent: the waveform at time $t$ is determined by an integral over its entire past. This is precisely the structural feature needed for a memory-like effect. In addition, the additional degree of freedom provided by the fractional order $\alpha$ may allow the nonlocal kernel to mimic certain nonlinear features and thus make such an outcome possible. For this reason, the frequency response of the fractional model \eqref{eq: Fractional 1+3 EFE w dimensional consistency} and the permanent offset in $h_{\mu\nu}$ are the key elements to answer the fundamental question of this work.

To understand the behavior of \eqref{eq: Fractional 1+3 EFE w dimensional consistency}, it is convenient to analyze the relation under the assumption of spherical symmetry in both the field and the source. Therefore, the system is reduced to $1+1$ $D$ form with spherical coordinates $(t,r)$. This symmetry reduction enables us to understand the qualitative behavior of the model, including the late-time behavior, without isolating radiative degrees of freedom in the Bondi sense, nor taking a null-infinity limit. This reduction is used to understand the hereditary behavior of the fractional model with fewer control parameters and does not account for the memory in GR, since it does not have any memory in spherically symmetric fields. After this reduction, \eqref{eq: Fractional 1+3 EFE w dimensional consistency} becomes
\begin{equation} \label{eq: Fractional 1+1 EFE w radial coordinate with bar}
      -\frac{\Gamma(2-\alpha)}{c\,t^{1-\alpha}}\partial_{t}^{\alpha}\Bigl[\frac{\Gamma(2-\alpha)}{c\,t^{1-\alpha}}\partial_{t}^{\alpha}\bar{u}(t,r)\Bigr]+\frac{1}{r}\partial^2_r(r\,\bar{u}(t,r))= -\bar{s}(t,r),
  \qquad 0<\alpha<1,
\end{equation}
where $\bar{u}(t,r)$ and $\bar{s}(t,r)$ are arbitrary well-behaved fields for $r\in(0,\infty)$ and $t\in(0^+,\infty)$. To solve the relation numerically, it is convenient to define the functions $u=u(t,r)$ and $s=s(t,r)$ such that $u(t,r):= r\bar{u}(t,r)$ and $s(t,r):= r\bar{s}(t,r)$. After employing the new functions, we set ($c=1$) and solve it numerically (the numerical scheme is described in Appendix B):

\begin{equation}\label{eq: 1+1 Fractional Wave Equation}
  \frac{\Gamma(2-\alpha)}{t^{1-\alpha}}\partial_{t}^{\alpha}
  \!\left[
    \frac{\Gamma(2-\alpha)}{t^{1-\alpha}}\partial_{t}^{\alpha} u(t,r)
  \right]-\partial_{r}^{2}u(t,r)
  = s(t,r),
  \qquad 0<\alpha<1.
\end{equation}
We impose Dirichlet boundaries as $u(t,10)=u(t,30)=0$ for $t>0$ on the interval $r\in[10,30]$ to avoid boundary reflections within the observation time. The initial conditions are taken as $u(0^+,r)=0$, $u_t[0^+,r)=0$, and we take $s(t,r)$ to be localized in space and time:
\begin{equation}\label{eq: Source for 1+1 fractional wave equation}
  s(t,r)=\exp\!\Bigl(-\frac{(r-r_0)^{2}}{2s_r^{2}}\Bigr)\exp\!\Bigl(-\frac{(t-t_0)^{2}}{2s_t^{2}}\Bigr)\sin\!\bigl(\omega(t-t_0)\bigr).
\end{equation}
where $\omega\in\{-6\pi,-18\pi,-24\pi\}$, $r_0=20.0$, $t_0=0.3$, $s_r=0.05$, $s_t=0.06$. Increasing $|\omega|$ results in higher-frequency forcing. For each $\omega$, we compare several $\alpha$ values at two observation points.

\begin{figure}[H]
  \centering
  \begin{subfigure}{0.49\textwidth}
    \centering
    \includegraphics[width=\linewidth]{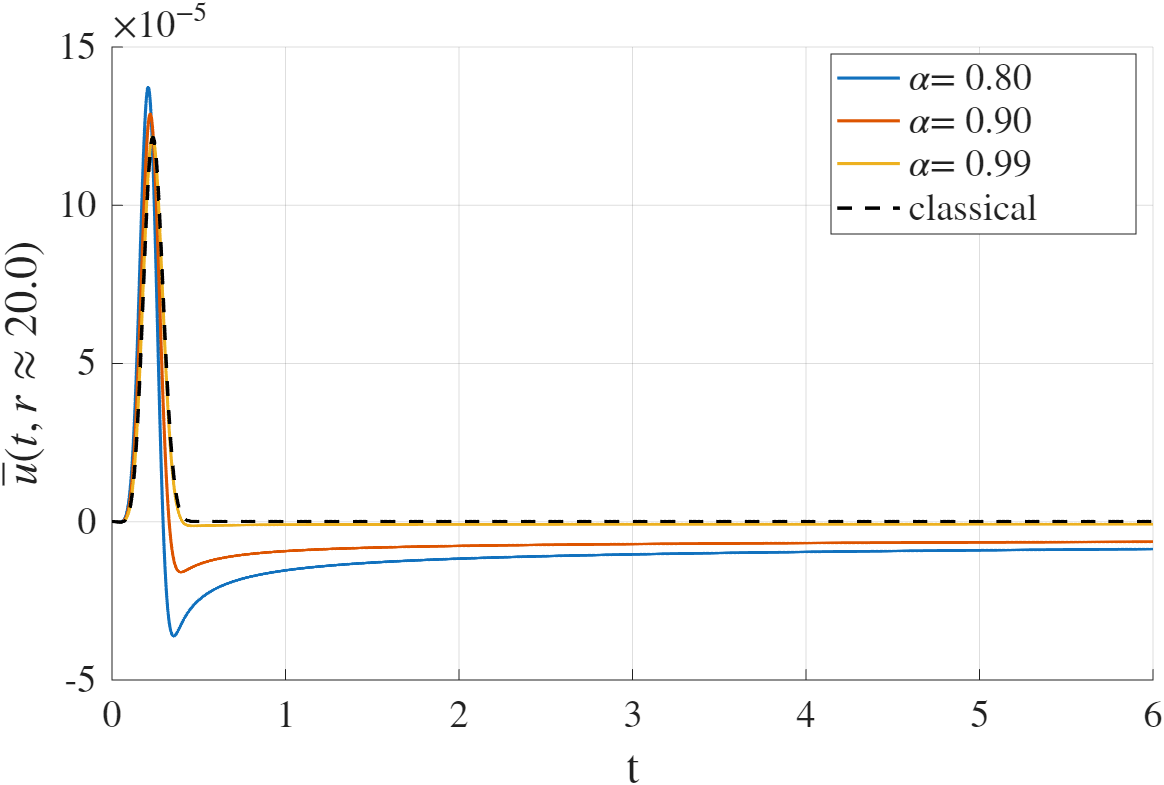}
    \caption{Observation location $r\approx 20.0$.}
    \label{fig:case1a}
  \end{subfigure}\hfill
  \begin{subfigure}{0.49\textwidth}
    \centering
    \includegraphics[width=\linewidth]{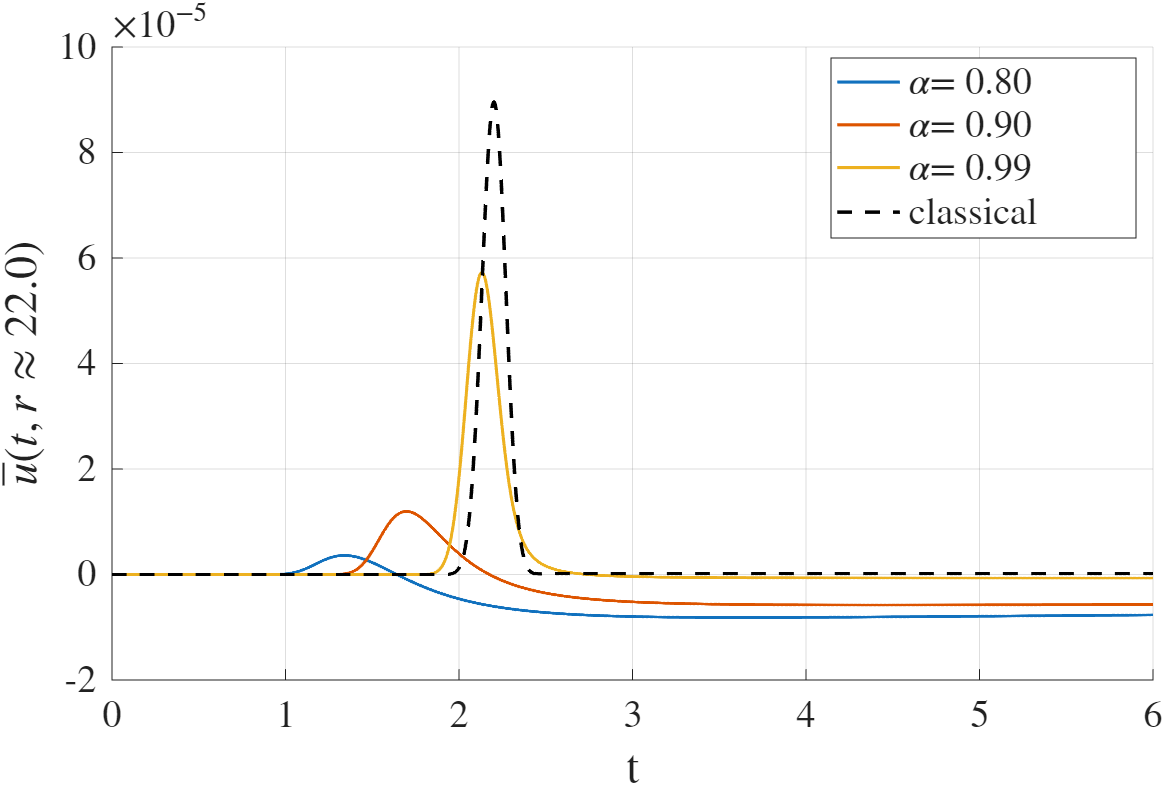}
    \caption{Observation location $r\approx 22.0$.}
    \label{fig:case1b}
  \end{subfigure}
  \caption{Time evolution of the field $\bar{u}(t,r)$ at two observation points (a) near the source ($r\approx20.0$) and (b) outside the source ($r\approx22.0$) for the source frequency $\omega=-6\pi$. The curves corresponds to several fractional orders $\alpha$ and the classical curve represents integer order solution of \eqref{eq: 1+1 Fractional Wave Equation}.}
  \label{fig:case1}
\end{figure}

\begin{figure}[H]
  \centering
  \begin{subfigure}{0.49\textwidth}
    \centering
    \includegraphics[width=\linewidth]{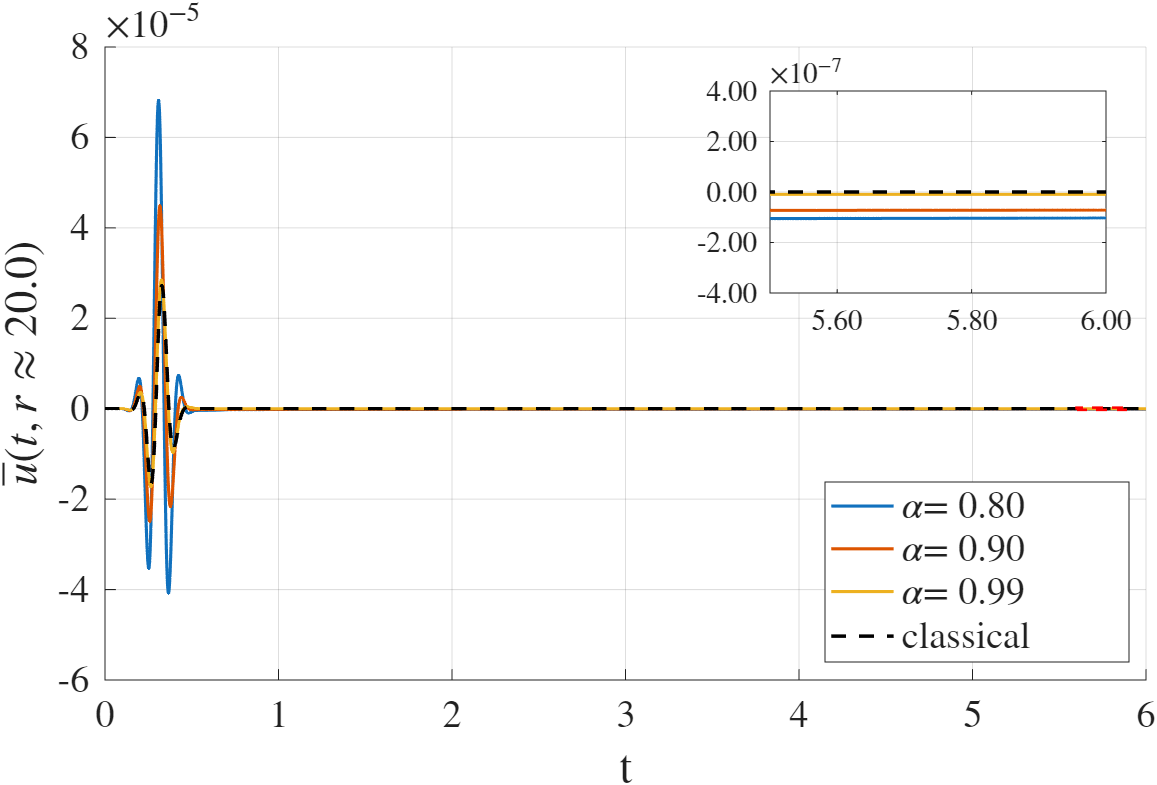}
    \caption{Observation location $r\approx 20.0$.}
    \label{fig:case2a}
  \end{subfigure}\hfill
  \begin{subfigure}{0.49\textwidth}
    \centering
    \includegraphics[width=\linewidth]{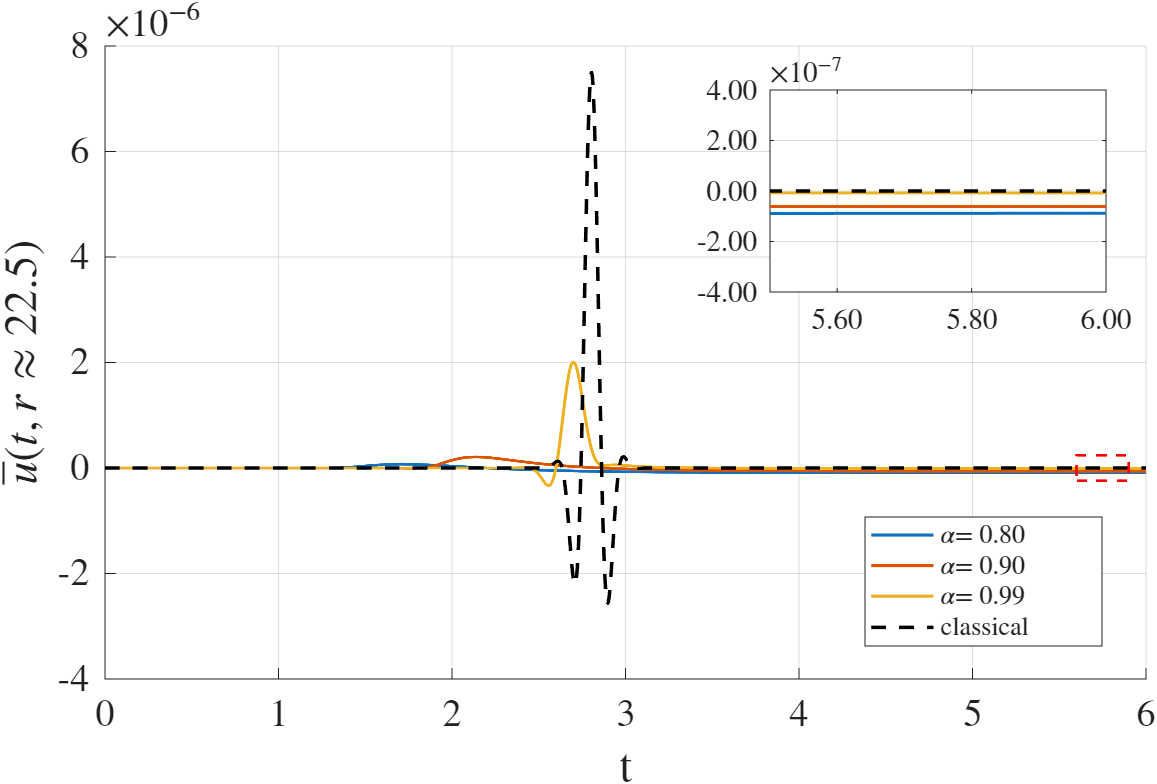}
    \caption{Observation location $r\approx 22.5$.}
    \label{fig:case2b}
  \end{subfigure}
  \caption{Time evolution of the field $\bar{u}(t,r)$ at two observation points (a) near the source ($r\approx20.0$) and (b) outside the source ($r\approx 22.5$) for the source frequency $\omega=-18\pi$. The top-right inset shows a zoomed-in view of the region enclosed by the red-dashed rectangle. The curves corresponds to several fractional orders $\alpha$ and the classical curve represents integer order solution of \eqref{eq: 1+1 Fractional Wave Equation}.}
  \label{fig:case2}
\end{figure}

\begin{figure}[H]
  \centering
  \begin{subfigure}{0.49\textwidth}
    \centering
    \includegraphics[width=\linewidth]{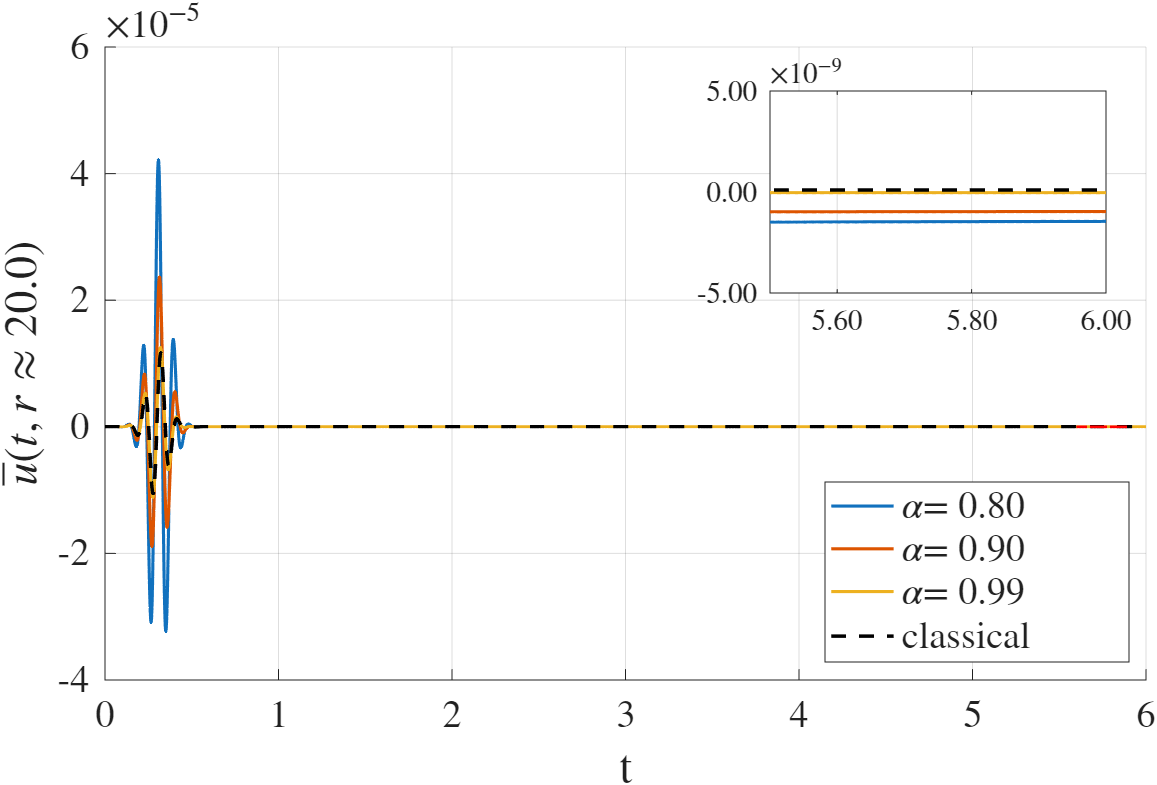}
    \caption{Observation location $r\approx 20.0$.}
    \label{fig:case3a}
  \end{subfigure}\hfill
  \begin{subfigure}{0.49\textwidth}
    \centering
    \includegraphics[width=\linewidth]{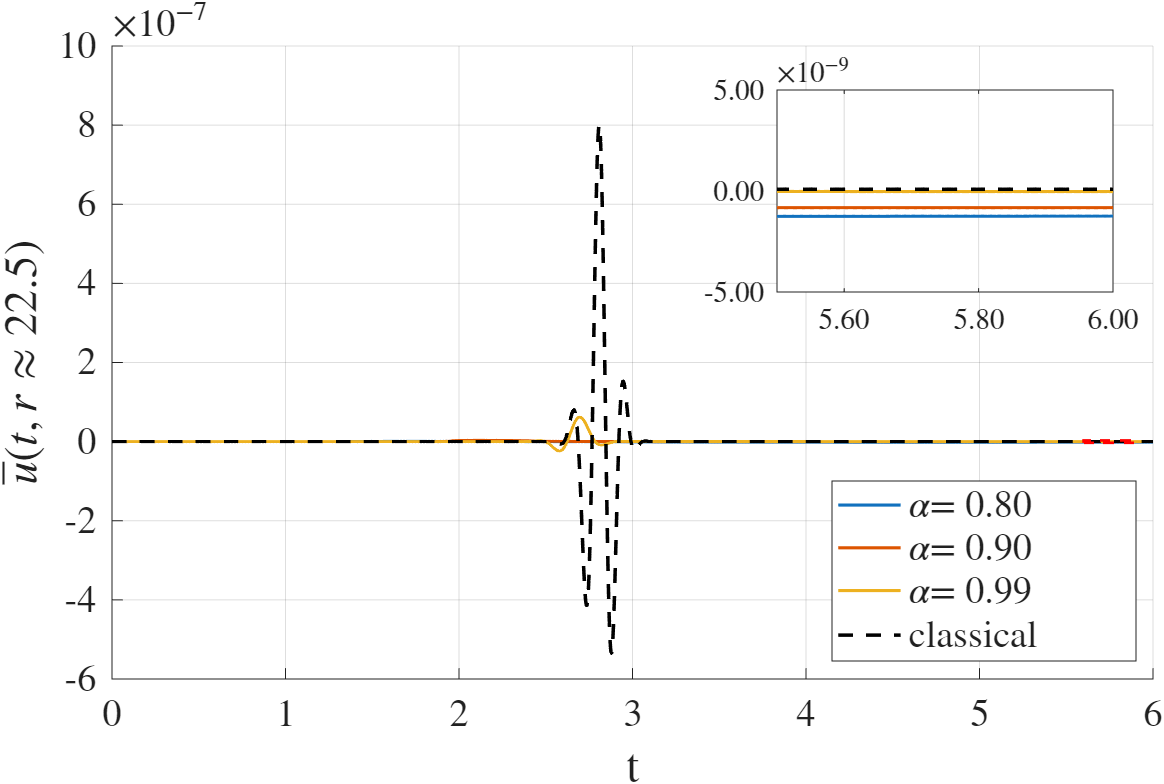}
    \caption{Observation location $r\approx 22.5$.}
    \label{fig:case3b}
  \end{subfigure}
  \caption{Time evolution of the field $\bar{u}(t,r)$ at two observation points (a) near the source ($r\approx20.0$) and (b) outside the source ($r\approx22.5$) for the source frequency $\omega=-24\pi$. The top-right inset shows a zoomed-in view of the region enclosed by the red-dashed rectangle. The curves corresponds to several fractional orders $\alpha$ and the classical curve represents integer order solution of \eqref{eq: 1+1 Fractional Wave Equation}.}
  \label{fig:case3}
\end{figure}

Let us note several observations. ($i$) In panels (a) of Figs.\,\ref{fig:case1}–\ref{fig:case3}, for $\alpha=0.99$, the field $\bar{u}(t,r\approx20.0)$ results are identical to the classical solution. However, in panels (b), the field $\bar{u}(t,r\approx 22.5)$ shows significant deviations from the classical solution. This pattern implies that the convergence of the sequential Caputo operator to $\partial_t^2$ as $\alpha \to 1^-$ requires closer values of $\alpha$ to $1$ for larger observation points.
($ii$) As illustrated in Figs.\,\ref{fig:case1}–\ref{fig:case3}, the sequential fractional model produces a small memory-like offset in the field $\bar{u}(t,r)$. The magnitude of this offset increases as the fractional order $\alpha$ decreases. However, the magnitude of this offset decreases as the source frequency $|\omega|$ increases. This trend runs opposite to the gravitational nonlinear memory predicted by GR, which grows with the total radiated energy flux (see \eqref{eq: the nonlinear memory}).
($iii$) Comparison of panels (a) and (b) in Figs.\,\ref{fig:case1}–\ref{fig:case3} shows that the apparent offset is slightly reduced at the larger observation distances $r$, indicating that the effect is localized near the source.
($iv$) In all cases, the simulations show that the solution decays to zero at late times, regardless of $\alpha$ or $\omega$. This confirms that the operator in \eqref{eq: 1+1 Fractional Wave Equation} behaves like a damped diffusion–wave system rather than producing a permanent displacement in the field.

Overall, the fractional model captures some hereditary, memory-like features, but it fails to reproduce the actual nonlinear memory effect of GR, which requires a permanent asymptotic offset. In additional simulations with different choices of the source parameters in \eqref{eq: Source for 1+1 fractional wave equation}, we observe the same qualitative trends. It is worth noting that varying the source parameter $t_0$, which alters the source initialization in \eqref{eq: Source for 1+1 fractional wave equation}, alters the field results $\bar{u}(t,r)$ for every $\alpha$. These results verify the broken time translation invariance of the chosen Caputo operator form. In particular, as $t_0 \to 0^+$ in \eqref{eq: Source for 1+1 fractional wave equation}, the offset in $\bar{u}(t,r)$ increases the amplitude; however, the field $\bar{u}(t,r)$ vanishes more rapidly. These results are consistent with the power-law weights in \eqref{eq: L1 scheme}.

For effective modeling, one may combine the classical and fractional operators as follows:
\begin{equation}\label{eq: Classical/Fractional wave equation combination}
      A\Box \bar{h}_{\mu\nu}+ B\left(-\frac{\Gamma(2-\alpha)}{c\,t^{1-\alpha}}\partial_{t}^{\alpha}\Bigl[\frac{\Gamma(2-\alpha)}{c\,t^{1-\alpha}}\partial_{t}^{\alpha}\Bigr]+\Delta\right)\bar{h}_{\mu\nu}=-\frac{16\pi G}{c^{4}}T_{\mu\nu}.
\end{equation}
With coefficients $A$ and $B$ satisfying $A+B=1$, we can allow more general coefficients; however, we have chosen their sum to be one below.

\subsection{Fractional quadrupole moment}
\label{Fractional Quadrupole Moment}

In Sec.\,\ref{Modification of the Linearized Einstein Field Equation}, the sequential time–fractional wave model shows a memory-like offset, but its solutions exhibit damped oscillation dynamics. To better preserve the wave nature of the linearized EFE, we suggest a hybrid formulation \eqref{eq: Classical/Fractional wave equation combination}. Nevertheless, for sufficiently large radiated energy flux, the model still fails to reproduce the qualitative features of the nonlinear memory. The numerical results do not clarify whether the qualitative behavior arises from the fractional wave equation or from the kernel in the Caputo derivative. In addition, requiring physically meaningful initial conditions restricts the possible set of admissible fractional derivatives. Therefore, we consider a naive fractionalization of the integer-order model to understand the behavior of the sequential Caputo fractional derivative. This approach has two advantages: there are no restrictions on the admissible fractional derivative, and linearity allows the superposition of possible fractionalizations. Hence, the model admits many alternative fractionalization choices, and one can evaluate the role of the kernel more directly. Furthermore, if this modification captures the nonlinear memory effect, it provides a new computational tool for calculating memory. To explore these possibilities, we begin by introducing a naive fractional modification of the quadrupole moment. In the far-zone and the slow motion limit \cite{Maggiore2008book},
\begin{equation}\label{eq: Quadrupole moment-classical}
  \bigl[h^{\mathrm{TT}}_{ij}(t,\boldsymbol{x})\bigr]_{\mathrm{quad}}'=\frac{2G}{c^{4} r}\Lambda_{ij,kl}(\hat{\boldsymbol n})\,\ddot{Q}_{kl}\!\left(t-\frac{r}{c}\right),
\end{equation}
where $\Lambda_{ij,kl}$ is the projection tensor and $r=|\boldsymbol{x}|$.

We consider the following fractionalized modification of the second time derivative acting on the source moment:
\begin{equation}\label{eq: Quadrupole Moment Fractional}
  \bigl[{}_{(\alpha)}h^{\mathrm{TT}}_{ij}(t,\boldsymbol{x})\bigr]_{\mathrm{quad}}= \frac{2G}{c^{4} r}\,\Lambda_{ij,kl}\!\bigl(\hat{\boldsymbol n}\bigr)\,\frac{\Gamma(2-\alpha)}{u^{1-\alpha}}\partial_{u}^{\alpha}\!\left[\frac{\Gamma(2-\alpha)}{u^{1-\alpha}}\,\partial_{u}^{\alpha} Q_{kl}(u)\right],\qquad u := t-\frac{r}{c},
\end{equation}
where $u$ is the retarded time and $0<\alpha<1$. As in \eqref{eq: Fractional 1+3 EFE w dimensional consistency}, this sequential Caputo form is dimensionally consistent; however, the modification does not preserve all tensorial/gauge properties of linearized gravity, similar to the model \eqref{eq: Fractional 1+3 EFE w dimensional consistency}. For effective modeling, we employ a linear combination of \eqref{eq: Quadrupole moment-classical} and \eqref{eq: Quadrupole Moment Fractional},
\begin{equation}\label{eq: Linear combination of Quadrupole moment classical/fractional}
  \bigl[h^{\mathrm{TT}}_{ij}\bigr]_{\mathrm{quad}} = A\,\bigl[{}_{(\alpha)}h^{\mathrm{TT}}_{ij}\bigr]_{\mathrm{quad}} + B\,\bigl[h^{\mathrm{TT}}_{ij}\bigr]_{\mathrm{quad}}',
\end{equation}
where $A, B$ are dimensionless weights chosen so that $A+B=1$; one can allow for more general coefficients similar to the effective model \eqref{eq: Classical/Fractional wave equation combination}. For weights, $A, B \in [0,1]$ is chosen as a restriction, but a larger subset can also be selected. In addition, the GR limit corresponds to $A=0$, $B=1$.

To investigate the offset response as a function of source frequency, we consider a toy model without a physical background. An artificial equal-mass binary with $m_1=m_2=1.4\,M_{\odot}$ in the $xy$-plane and center of mass (CM) at the origin. Let the relative separation be
\begin{equation}
  \boldsymbol{r}(t) = a(t)[\cos\phi(t),\,\sin\phi(t),\,0], 
  \qquad
  a(t)=a_0\exp\left[-\frac{(t-t_0)^2}{2s_t^{2}}\right],
\end{equation}
with $a_0=0.2\,\mathrm{km}$, $t_0=20\,\mathrm{s}$, $s_t=7.15\,\mathrm{s}$, and $\phi(t)=\omega_0 t$ where $\omega_0=2\pi f_0$, $f_0\in\{0.2,0.4,0.6\}\,\mathrm{Hz}$. The component positions (CM frame) are
\begin{equation}
  \boldsymbol{x}_1(t)=\frac{m_2}{M}\,\boldsymbol{r}(t), 
  \qquad
  \boldsymbol{x}_2(t)=-\,\frac{m_1}{M}\,\boldsymbol{r}(t),
  \qquad M=m_1+m_2,
\end{equation}
and we observe along the $\hat{\boldsymbol z}$ axis at a distance $r=0.0065 \,\mathrm{pc}$.

\begin{figure}[H]
  \centering
  \begin{subfigure}{0.49\textwidth}
    \centering
    \includegraphics[width=\linewidth]{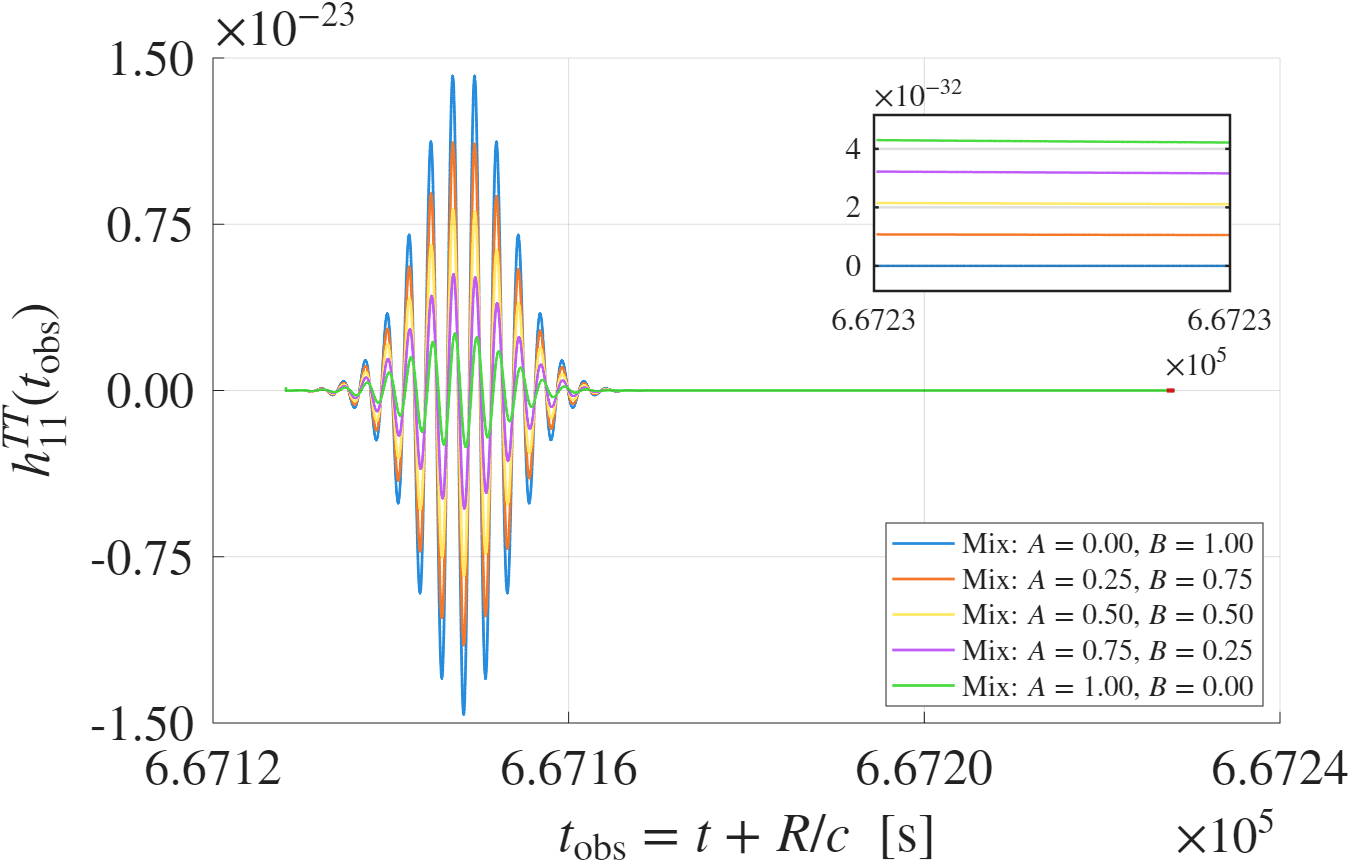}
    \caption{$\alpha=0.8$.}
    \label{fig:RHS_case1_a08}
  \end{subfigure}\hfill
  \begin{subfigure}{0.49\textwidth}
    \centering
    \includegraphics[width=\linewidth]{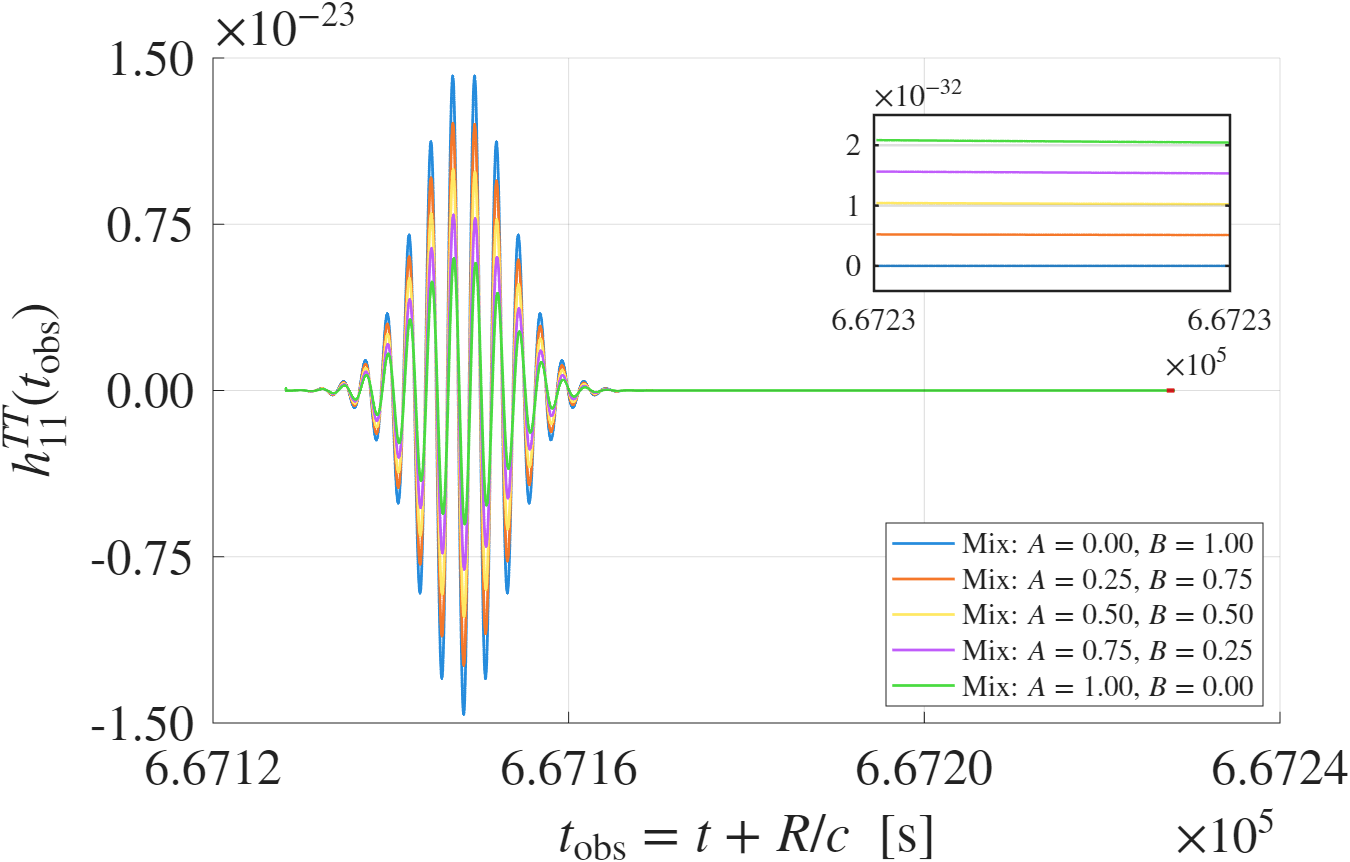}
    \caption{$\alpha=0.9$.}
    \label{fig:RHS_case1_a09}
  \end{subfigure}
  \caption{Time evolution of the transverse-traceless waveform $h_{11}^{\mathrm{TT}}$ for an equal mass artificial binary with orbital frequency $f_0=0.2\,\mathrm{Hz}$, observed along the $z$-axis. The two panels compare the fractional model for (a) $\alpha=0.8$ and (b) $\alpha=0.9$. In panels (a) and (b), the bottom-right inset shows the chosen A and B values in \eqref{eq: Linear combination of Quadrupole moment classical/fractional}, and the top-right inset shows the zoomed-in view of the region enclosed by the red point.}
  \label{fig:RHS_case1}
\end{figure}

\begin{figure}[H]
  \centering
  \begin{subfigure}{0.49\textwidth}
    \centering
    \includegraphics[width=\linewidth]{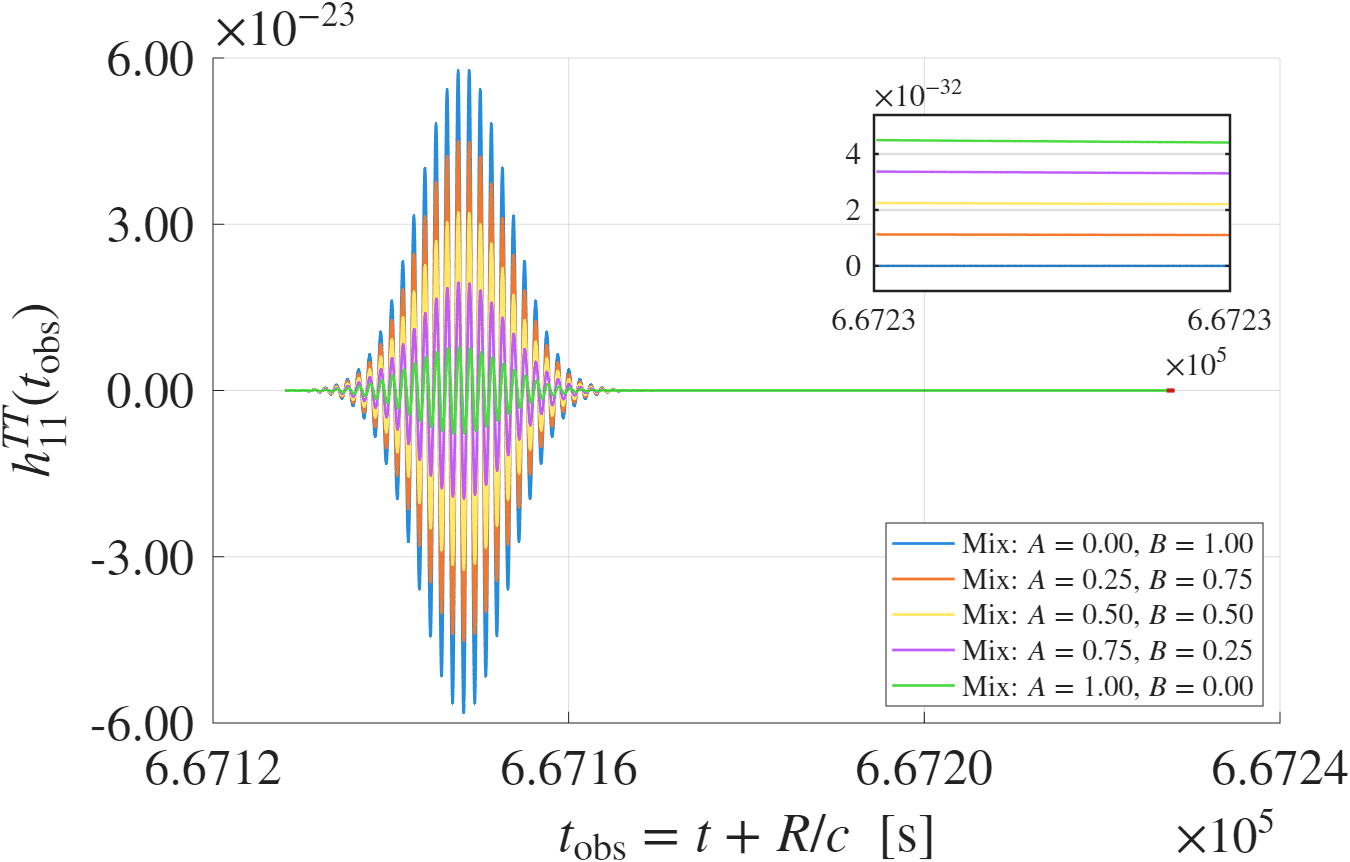}
    \caption{$\alpha=0.8$.}
    \label{fig:RHS_case2_a08}
  \end{subfigure}\hfill
  \begin{subfigure}{0.49\textwidth}
    \centering
    \includegraphics[width=\linewidth]{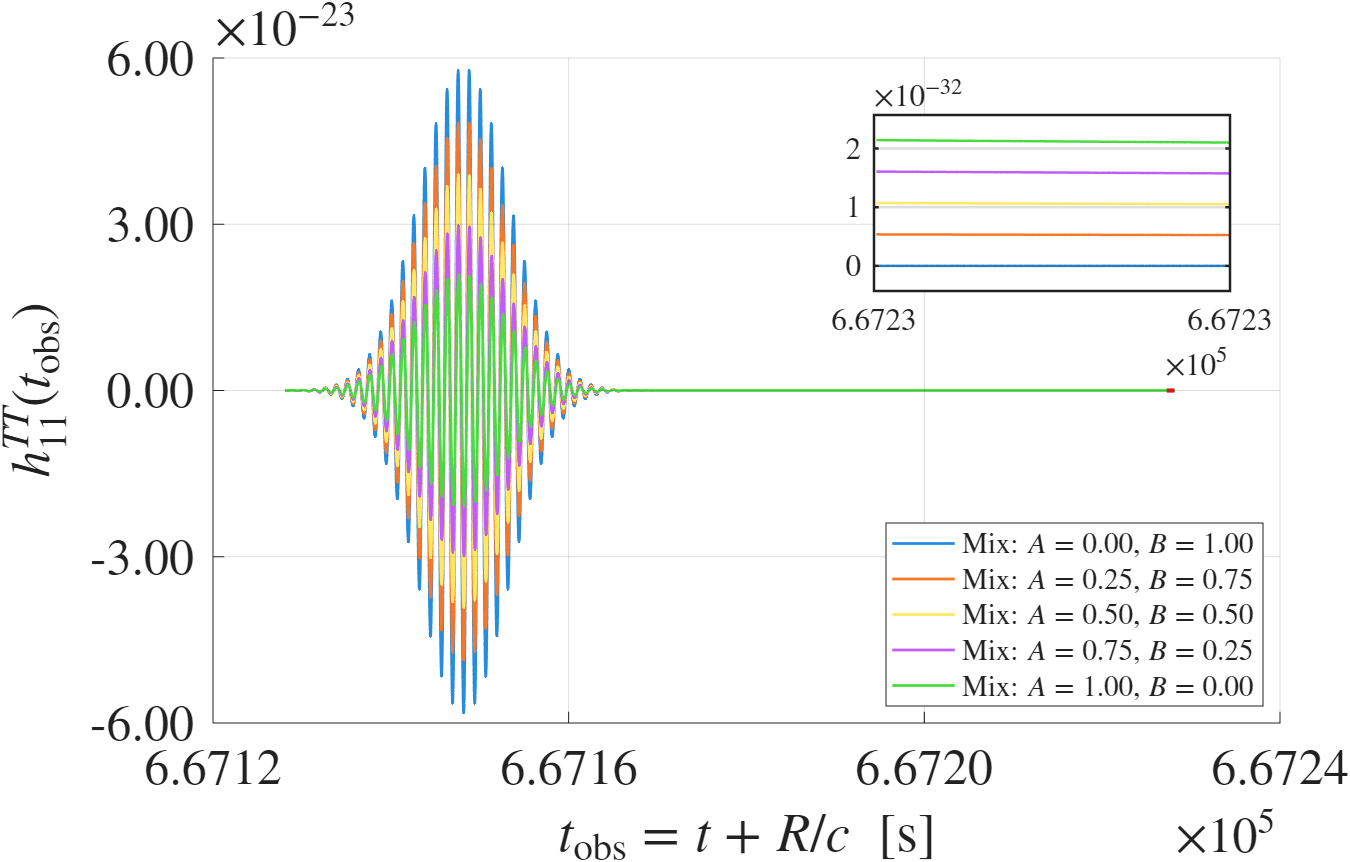}
    \caption{$\alpha=0.9$.}
    \label{fig:RHS_case2_a09}
  \end{subfigure}
  \caption{Time evolution of the transverse-traceless waveform $h_{11}^{\mathrm{TT}}$ for an equal mass artificial binary with orbital frequency $f_0=0.4\,\mathrm{Hz}$, observed along the $z$-axis. The two panels compare the fractional model for (a) $\alpha=0.8$ and (b) $\alpha=0.9$. In panels (a) and (b), the bottom-right inset shows the chosen A and B values in \eqref{eq: Linear combination of Quadrupole moment classical/fractional}, and the top-right inset shows the zoomed-in view of the region enclosed by the red point.}
  \label{fig:RHS_case2}
\end{figure}

\begin{figure}[H]
  \centering
  \begin{subfigure}{0.49\textwidth}
    \centering
    \includegraphics[width=\linewidth]{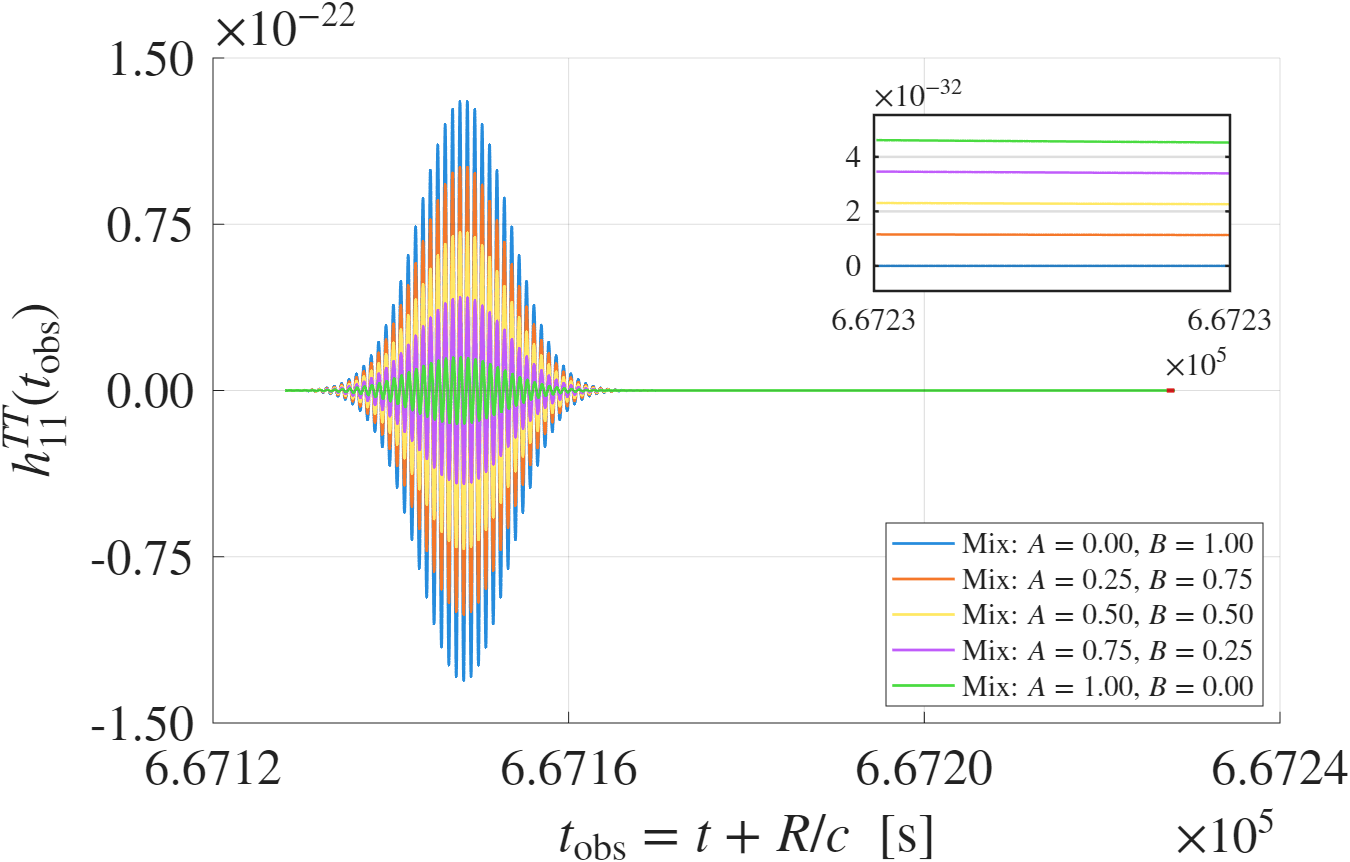}
    \caption{$\alpha=0.8$.}
    \label{fig:RHS_case3_a08}
  \end{subfigure}\hfill
  \begin{subfigure}{0.49\textwidth}
    \centering
    \includegraphics[width=\linewidth]{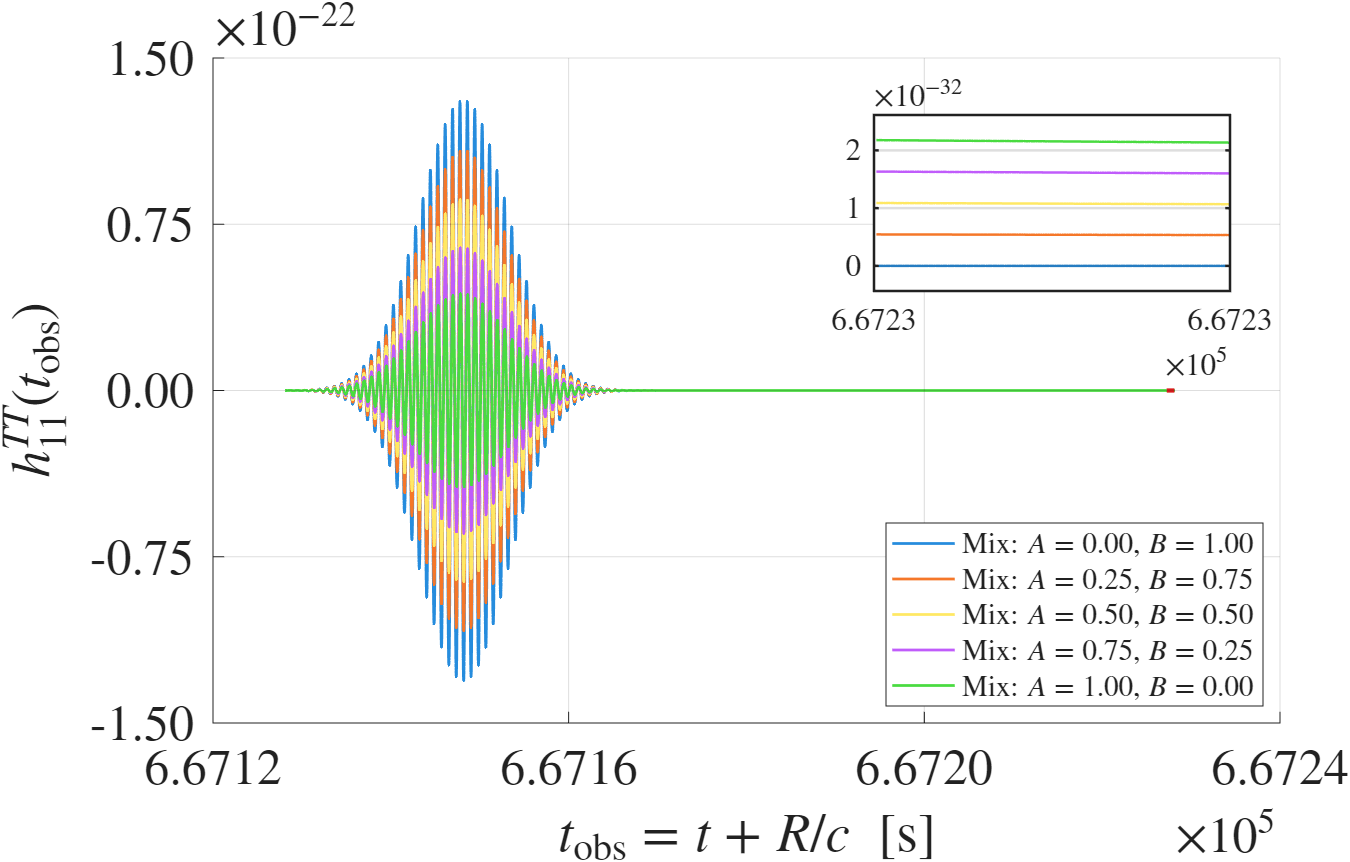}
    \caption{$\alpha=0.9$.}
    \label{fig:RHS_case3_a09}
  \end{subfigure}
  \caption{Time evolution of the transverse-traceless waveform $h_{11}^{\mathrm{TT}}$ for an equal mass artificial binary with orbital frequency $f_0=0.6\,\mathrm{Hz}$, observed along the $z$-axis. The two panels compare the fractional model for (a) $\alpha=0.8$ and (b) $\alpha=0.9$. In panels (a) and (b), the bottom-right inset shows the chosen A and B values in \eqref{eq: Linear combination of Quadrupole moment classical/fractional}, and the top-right inset shows the zoomed-in view of the region enclosed by the red point.}
  \label{fig:RHS_case3}
\end{figure}

As shown in Figs.\,\ref{fig:RHS_case1}–\ref{fig:RHS_case3}, the apparent memory, measured by the late–time offset of $h_{11}^{\mathrm{TT}}$, increases slightly as the binary orbital frequency $f_0$ increases and as the fractional order $\alpha$ decreases. This trend is visible by comparing different simulations: for example, in Fig.\,\ref{fig:RHS_case1}, the offset is larger in (a) ($\alpha=0.8$) than in (b) ($\alpha=0.9$), and the same behavior persists at higher frequencies in Figs.\,\ref{fig:RHS_case2} and \ref{fig:RHS_case3}. These results suggest that the contribution of the fractional operator depends both on $\alpha$ and the source frequency. Nevertheless, in all numerical simulations, the waveform decays incrementally to zero as $t \to \infty$, indicating that the fractionalization of the quadrupole moment cannot produce a permanent displacement in the detectors. Thus, it still fails to capture the nonlinear memory predicted by GR, which requires a permanent asymptotic offset related to the radiated energy flux from GWs.

\section{Late-Time Behavior of Fractional Wave Model and Gravitational Memory}\label{Late-Time Behavior of Fractional Wave Model and Gravitational Memory}
The purpose of this section is not to model unbound radiative flux but to establish that fractional time operators without an explicit flux-balance mechanism necessarily suppress late-time offsets. The original motivation for this model was to test whether a fractional time derivative operator could support a permanent offset in the metric perturbation after the passage of a localized gravitational burst. The results below show that, under a set of mathematically precise assumptions, the fractional model does \emph{not} support a permanent memory effect.

\subsection{Fractional Wave Model}
Consider the field equation
\begin{equation}\label{eq: Fractional 1+n EFE w simplied notation}
 -t^{\alpha-1} \partial_t^{\alpha}\!\left(t^{\alpha-1}\partial_t^{\alpha}\bar{h}_{\mu\nu}\right) + \Delta \bar{h}_{\mu\nu} = -T_{\mu\nu},
\end{equation}
with $0<\alpha<1$. The parameter $\alpha$ defines a fractional time differentiation structure \eqref{eq: Caputo fractional derivative definition} and generalizes the standard wave equation. We assume spatial flatness and standard asymptotic decay in the domain $(t,x)\in[0^+,\infty)\times\mathbb{R}^n$. Note that $0^+$ denotes the {\it lower terminal} of the Caputo fractional derivative in \eqref{eq: Caputo fractional derivative definition} rather than a limit.

We now analyze the late-time behavior of \eqref{eq: Fractional 1+n EFE w simplied notation} after the source becomes inactive. Suppose that the stress-energy tensor is compactly supported in both time and space, so that for some finite $t_f$
\begin{equation}
    T_{\mu\nu}(t,x)=0 
    \quad \forall\, x\in\mathbb{R}^n,\quad t>t_f.
\end{equation}
Then, for $t>t_f$, \eqref{eq: Fractional 1+n EFE w simplied notation} reduces to
\begin{equation} \label{eq: Fractional 1+n w homogenous}
 -t^{\alpha-1} \partial_t^{\alpha}\!\left(t^{\alpha-1}\partial_t^{\alpha}\bar{h}_{\mu\nu}\right)
 + \Delta \bar{h}_{\mu\nu} = 0.
\end{equation}
We additionally impose the initial conditions
\begin{equation} \label{eq: initial conditions}
\bar{h}_{\mu\nu}(0^+,x)=0,
\qquad
\partial_t \bar{h}_{\mu\nu}(0^+,x)=0,
\end{equation}
which corresponds to no pre-existing radiation before the wave pulse.

Let us define the function
\begin{equation}
 \bar{g}^\alpha_{\mu\nu}(t,x)
 :=
 \partial_t^\alpha \!\left(
 t^{\alpha-1}\partial_t^\alpha \bar{h}_{\mu\nu}
 \right),
 \label{eq:gbarDef}
\end{equation}
where the superscript $\alpha$ labels the fractional order and is not a tensor index. Then equation \eqref{eq: Fractional 1+n w homogenous} becomes
\begin{equation}
  -t^{\alpha-1}\bar{g}^\alpha_{\mu\nu}
  +\Delta \bar{h}_{\mu\nu}=0.
  \label{eq:equationrewrite}
\end{equation}
The strategy is to show that the quantity $t^{\alpha-1}\bar{g}^\alpha_{\mu\nu}$ vanishes at late times. If so, the system reduces to the Laplace equation. In the rest of this section, we show that $\bar{g}^\alpha_{\mu\nu}$ is uniformly bounded when there are certain restrictions on the field $\bar{h}_{\mu\nu}$.

We impose the following admissible physical hypothesis:\\[4pt]
\noindent
\textbf{Assumption A.} \emph{
All time derivatives of $\bar{h}_{\mu\nu}$ are temporally localized for any observation location and are bounded for all $(t,x)\in [0^+,\infty) \times\mathbb{R}^n$. In particular, for each $k\in\mathbb{N}^+$, there exists $M_k>0$ such that $|\partial_t^k\bar{h}_{\mu\nu}(t,x)|<M_k$ holds for all $(t,x)$, and for $\forall x\in {\mathbb{R}^n}$, there exists $\tau'\in\mathbb{R^+}$ such that $\partial_t^k\bar{h}_{\mu\nu}(t,x)=0$ is true for $t>\tau'.
$
}\\[4pt]
This corresponds to a finite total emitted radiation and a finite energy flux, which are physically reasonable for burst-like systems.

Here we show that, under \emph{Assumption A}, $\bar{g}^\alpha_{\mu\nu}$ is uniformly bounded for all $\mu,\nu$ when $0<\alpha<1$. First, we write $\bar{g}^\alpha_{\mu\nu}$ explicitly, and for notational simplicity, we omit the constant prefactors involving $\Gamma(1-\alpha)$, since they do not affect the boundedness.
\begin{equation}\label{eq: first bar g_munu explicitly}
    \bar{g}^\alpha_{\mu\nu}(t,x)
    =
    \int_{0^+}^td\tau(t-\tau)^{-\alpha}\partial_\tau \left(\tau^{-1+\alpha}\int_{0^+}^\tau d\xi(\tau-\xi)^{-\alpha} \partial_\xi\bar{h}_{\mu\nu}(\xi,x)\right).
\end{equation}
Applying the $\tau$-derivative using the product rule, we obtain
\begin{equation}
    \bar{g}^\alpha_{\mu\nu}
    =
    \int_{0^+}^t d\tau \frac{(-1+\alpha)\tau^{-2+\alpha}}{(t-\tau)^{\alpha}}\int_{0^+}^\tau d\xi \frac{\partial_\xi\bar{h}_{\mu\nu}}{(\tau-\xi)^{\alpha}}
    +
    \int_{0^+}^t d\tau \frac{\tau^{-1+\alpha} }{(t-\tau)^{\alpha}}\partial_\tau\left (\int_{0^+}^\tau d\xi \frac{\partial_\xi\bar{h}_{\mu\nu}}{(\tau-\xi)^{\alpha}}\right).
\end{equation}
To handle the weak singularity in the kernel $(\tau-\xi)^{-\alpha}$ for $0<\alpha<1$, we first integrate by parts with respect to $\xi$

\begin{equation}
    \int_{0^+}^\tau d\xi \frac{1}{(\tau-\xi)^\alpha}\partial_\xi\bar{h}_{\mu\nu}
    =
    -\frac{(\tau-\xi)^{1-\alpha}}{1-\alpha}\partial_\xi\bar{h}_{\mu\nu}\bigg|_{0^+=\xi}^{\tau}
    +
    \int_{0^+}^\tau d\xi \frac{(\tau-\xi)^{1-\alpha}}{1-\alpha}\partial_\xi^2\bar{h}_{\mu\nu}.
\end{equation}
The boundary contribution at $\xi=\tau$ vanishes since $(\tau-\tau)^{1-\alpha}=0$. Moreover, using the initial conditions \eqref{eq: initial conditions}, the $\xi=0^+$ boundary term also vanishes. Differentiating with respect to $\tau$ by applying the Leibniz rule yields
\begin{equation}
    \partial_\tau \left(\int_{0^+}^\tau d\xi \frac{(\tau-\xi)^{1-\alpha}}{1-\alpha}\partial_\xi^2\bar{h}_{\mu\nu}\right)
    =
    \int_{0^+}^\tau d\xi (\tau-\xi)^{-\alpha}\partial_\xi^2\bar{h}_{\mu\nu},
\end{equation}
where the upper boundary term vanishes since $(\tau-\tau)^{1-\alpha}=0$. Then \eqref{eq: first bar g_munu explicitly} becomes

\begin{equation} \label{eq: bar g_munu after leibniz}
    \bar{g}^\alpha_{\mu\nu}
    =
    \int_{0^+}^t d\tau \frac{(-1+\alpha)\tau^{-2+\alpha}}{(t-\tau)^{\alpha}}\int_{0^+}^\tau d\xi \frac{\partial_\xi\bar{h}_{\mu\nu}}{(\tau-\xi)^{\alpha}}
    +
    \int_{0^+}^t d\tau \frac{\tau^{-1+\alpha} }{(t-\tau)^{\alpha}}\int_{0^+}^\tau d\xi \frac{\partial^2_\xi\bar{h}_{\mu\nu}}{(\tau-\xi)^{\alpha}}.
\end{equation}
By \emph{Assumption A}, for each fixed $x\in\mathbb R^n$ there exists $\tau'\in\mathbb{R}^+$ such that for all $k\in\{1,2\}$, $\partial_t^k\bar h_{\mu\nu}(t,x)=0$ when $t>\tau'.$ Then, for the inner integrals in \eqref{eq: bar g_munu after leibniz}
\begin{equation}
    \int^\tau_{0^+}d\xi\frac{\partial^k_\xi\bar{h}_{\mu\nu}}{(\tau-\xi)^{\alpha}}
    = 
    \int^{\tau'}_{0^+}d\xi\frac{\partial^k_\xi\bar{h}_{\mu\nu}}{(\tau-\xi)^{\alpha}}
    +
    \int^\tau_{\tau'}d\xi\frac{\partial^k_\xi\bar{h}_{\mu\nu}}{(\tau-\xi)^{\alpha}},
\end{equation}
the second term vanishes identically. Therefore, in the inner integrals, we may replace the upper limit $\tau$ by $\tau'$. Splitting the outer $\tau$-integration for $\tau<\tau'$ and $\tau>\tau'$, then \eqref{eq: bar g_munu after leibniz} becomes

\begin{align}
    \bar{g}^\alpha_{\mu\nu}
    =
    &\int_{\tau'}^t d\tau \frac{(-1+\alpha)\tau^{-2+\alpha}}{(t-\tau)^{\alpha}}\int_{0^+}^{\tau'} d\xi \frac{\partial_\xi\bar{h}_{\mu\nu}}{(\tau-\xi)^{\alpha}}
    +
    \int_{\tau'}^t d\tau \frac{\tau^{-1+\alpha} }{(t-\tau)^{\alpha}}\int_{0^+}^{\tau'} d\xi \frac{\partial^2_\xi\bar{h}_{\mu\nu}}{(\tau-\xi)^{\alpha}} \nonumber \\
    &+
    \int_{0^+}^{\tau'} d\tau \frac{(-1+\alpha)\tau^{-2+\alpha}}{(t-\tau)^{\alpha}}\int_{0^+}^{\tau} d\xi \frac{\partial_\xi\bar{h}_{\mu\nu}}{(\tau-\xi)^{\alpha}}
    +
    \int_{0^+}^{\tau'} d\tau \frac{\tau^{-1+\alpha} }{(t-\tau)^{\alpha}}\int_{0^+}^{\tau} d\xi \frac{\partial^2_\xi\bar{h}_{\mu\nu}}{(\tau-\xi)^{\alpha}}.
\end{align}
Now we use the boundedness of time derivatives in \emph{Assumption A}. Let $M:=\max\{M_1,M_2\}$ such that $|\partial_t \bar h_{\mu\nu}(t,x)|\le M$ and $|\partial_t^2 \bar h_{\mu\nu}(t,x)|\le M$. Then, we can put an upper bound for $\bar{g}^\alpha_{\mu\nu}$

\begin{align}\label{eq: definition of Balpha(t)}
    \frac{|\bar{g}^\alpha_{\mu\nu}|}{M}<B^\alpha(t):=&\int_{\tau'}^t d\tau \frac{(1-\alpha)\tau^{-2+\alpha}}{(t-\tau)^{\alpha}}\int_{0^+}^{\tau'} d\xi \frac{1}{(\tau-\xi)^{\alpha}}
    +
    \int_{\tau'}^t d\tau \frac{\tau^{-1+\alpha} }{(t-\tau)^{\alpha}}\int_{0^+}^{\tau'} d\xi \frac{1}{(\tau-\xi)^{\alpha}} \nonumber \\
    &+
    \int_{0^+}^{\tau'} d\tau \frac{(1-\alpha)\tau^{-2+\alpha}}{(t-\tau)^{\alpha}}\int_{0^+}^{\tau} d\xi \frac{1}{(\tau-\xi)^{\alpha}}
    +
    \int_{0^+}^{\tau'} d\tau \frac{\tau^{-1+\alpha} }{(t-\tau)^{\alpha}}\int_{0^+}^{\tau} d\xi \frac{1}{(\tau-\xi)^{\alpha}}.
\end{align}
Note that $-1+\alpha<0$ for $0<\alpha<1$. Thus, when taking an upper bound, we use
$|-1+\alpha|=1-\alpha$, so that all prefactors become positive. From this point on, we bound $|\bar g^\alpha_{\mu\nu}|$ by a positive function $B^\alpha(t)$. Therefore, our task reduces to analyze $B^\alpha(t)$. If we can show that $B^\alpha(t)$ is uniformly bounded under \emph{Assumption A}, then it follows that $\bar g^\alpha_{\mu\nu}$ is also uniformly bounded. To proceed, we first evaluate the common inner integral in the definition of $B^\alpha(t)$ in \eqref{eq: definition of Balpha(t)}

\begin{equation} \label{eq: explicit solution of the integral}
\int_{0^+}^{\tau'} d\xi(\tau-\xi)^{-\alpha}
=
-\frac{(\tau-\xi)^{1-\alpha}}{1-\alpha}\Big|_{\xi=0^+}^{\tau'}
=
\frac{\tau^{1-\alpha}-(\tau-\tau')^{1-\alpha}}{1-\alpha},
\end{equation}
for $\tau\ge \tau'$. Then, $B^\alpha(t)$ becomes
\begin{align}\label{eq: Balpha last}
B^\alpha(t)
=
&\int_{\tau'}^{t} d\tau \frac{1-(1-\frac{\tau'}{\tau})^{1-\alpha}}{(t-\tau)^\alpha \tau}
+
\int_{\tau'}^{t} d\tau \frac{1-(1-\frac{\tau'}{\tau})^{1-\alpha}}{(1-\alpha) (t-\tau)^\alpha} \nonumber \\
&+
\int_{0^+}^{\tau'} d\tau \frac{1}{(t-\tau)^\alpha \tau}
+
\int_{0^+}^{\tau'} d\tau \frac{1}{(1-\alpha)(t-\tau)^\alpha}.
\end{align}
The final form of $B^\alpha(t)$ in \eqref{eq: Balpha last} is not feasible for a direct analysis. Therefore, we introduce a more convenient upper bound by using the following inequality. For $0<\alpha<1$ and $\tau>\tau'$, we have

\begin{equation}
    (1-\frac{\tau'}{\tau})^{1-\alpha} > (1-\frac{\tau'}{\tau}),
\end{equation}
and therefore
\begin{equation}\label{eq: inequality}
1-(1-\frac{\tau'}{\tau})^{1-\alpha}
<
1-(1-\frac{\tau'}{\tau})
=
\frac{\tau'}{\tau}.
\end{equation}
By applying \eqref{eq: inequality}, we can bound $B^\alpha(t)$ from above by $\grave{B}^\alpha(t)$ as follows

\begin{align}\label{eq: definition graveB}
    \grave{B}^\alpha(t)
    :=&
    \int_{\tau'}^t d\tau \frac{{\tau'}}{(t-\tau)^{\alpha}\tau^2}
    +
    \int_{\tau'}^t d\tau \frac{{\tau'}}{(1-\alpha)(t-\tau)^{\alpha}\tau} \nonumber \\
    &+
    \int_{0^+}^{\tau'} d\tau \frac{1}{(t-\tau)^{\alpha}\tau}
    +
    \int_{0^+}^{\tau'} d\tau \frac{1}{(1-\alpha)(t-\tau)^{\alpha}}.
\end{align}
We remove the explicit $t$-dependence in the integrand via the change of variables $u=\frac{\tau}{t}$. We keep the last integral unchanged, since it can be evaluated by \eqref{eq: explicit solution of the integral}

\begin{align}
    \grave{B}^\alpha(t)
    =&
    \frac{\tau'}{t^{1+\alpha}}\int_{\frac{\tau'}{t}}^{1} du \frac{1}{(1-u)^{\alpha}u^2}
    +
    \frac{\tau'}{(1-\alpha)t^\alpha}\int_{\frac{\tau'}{t}}^1 du \frac{1}{(1-u)^{\alpha}u} \nonumber \\
    &+
    \frac{1}{t^\alpha}\int_{0^+}^{\frac{\tau'}{t}} du \frac{1}{(1-u)^{\alpha}u}
    +
    \frac{1}{1-\alpha}\int_{0^+}^{\tau'} d\tau \frac{1}{(t-\tau)^{\alpha}}.
\end{align}
For the first two integrals, we split them by introducing a constant $\lambda$ such that $1>\lambda>\frac{\tau'}{t}$ then

\begin{align} \label{eq: graveB last}
    \grave{B}^\alpha(t)
    =&
    \frac{\tau'}{t^{1+\alpha}}\int_{\frac{\tau'}{t}}^{\lambda} du \frac{1}{(1-u)^{\alpha}u^2}
    +
    \frac{\tau'}{t^{1+\alpha}}\int_{\lambda}^{1} du \frac{1}{(1-u)^{\alpha}u^2} \nonumber \\
    &+
    \frac{\tau'}{(1-\alpha)t^\alpha}\int_{\frac{\tau'}{t}}^\lambda du \frac{1}{(1-u)^{\alpha}u}
    +
    \frac{\tau'}{(1-\alpha)t^\alpha}\int_{\lambda}^1 du \frac{1}{(1-u)^{\alpha}u} \nonumber \\
    &+
    \frac{1}{t^\alpha}\int_{0^+}^{\frac{\tau'}{t}} du \frac{1}{(1-u)^{\alpha}u}
    +
    \frac{1}{1-\alpha}\int_{0^+}^{\tau'} d\tau \frac{1}{(t-\tau)^{\alpha}}.
\end{align}
We then bound $\grave{B}^\alpha(t)$ by applying separate bounds for each integrand.
\begin{enumerate}
    \item For $u\in[\lambda,1]$ we have $u \ge \lambda$; therefore, $u^{-1}\le \lambda^{-1}$ and $u^{-2}\le \lambda^{-2}$.
    \item For $u\in[\tau'/t,\lambda]$ we have $u\le \lambda \quad \Rightarrow \quad 1-u\ge 1-\lambda$, and therefore $(1-u)^{-\alpha}\le (1-\lambda)^{-\alpha}$.
\end{enumerate}
Substituting these bounds for each term in \eqref{eq: graveB last}, we define $\bar{B}^\alpha(t)$ so that $\grave{B}^\alpha(t) \le \bar{B}^\alpha(t)$ by

\begin{align} \label{eq: barB definition}
    \bar{B}^\alpha(t)
    :=&
    \frac{\tau'}{t^{1+\alpha}(1-\lambda)^\alpha}\int_{\frac{\tau'}{t}}^{\lambda} du \frac{1}{u^2}
    +
    \frac{\tau'}{t^{1+\alpha}\lambda^2}\int_{\lambda}^{1} du \frac{1}{(1-u)^{\alpha}}
    +
    \frac{\tau'}{(1-\alpha)t^\alpha(1-\lambda)^\alpha}\int_{\frac{\tau'}{t}}^\lambda du \frac{1}{u} \nonumber \\
    &+
    \frac{\tau'}{(1-\alpha)t^\alpha \lambda}\int_{\lambda}^1 du \frac{1}{(1-u)^{\alpha}}
    +
    \frac{1}{t^\alpha}\int_{0^+}^{\frac{\tau'}{t}} du \frac{1}{(1-u)^{\alpha}u}
    +
    \frac{1}{1-\alpha}\int_{0^+}^{\tau'} d\tau \frac{1}{(t-\tau)^{\alpha}}.
\end{align}
After solving the integrals in \eqref{eq: barB definition}, we obtain

\begin{align} \label{eq: last barB}
    \bar{B}^\alpha(t)
    =&
    \frac{\tau'}{t^{1+\alpha}(1-\lambda)^\alpha}(-\lambda^{-1}+\frac{t}{\tau'})
    +
    \frac{\tau'}{t^{1+\alpha}\lambda^2}\frac{(1-\lambda)^{1-\alpha}}{1-\alpha} 
    +
    \frac{\tau'}{(1-\alpha)t^\alpha(1-\lambda)^\alpha}ln(\frac{\lambda t}{\tau'}) \nonumber \\
    &+
    \frac{\tau'}{(1-\alpha)t^\alpha \lambda}\frac{(1-\lambda)^{1-\alpha}}{1-\alpha}
    +
    \frac{t^{1-\alpha}-(t-\tau')^{1-\alpha}}{(1-\alpha)^2}
    +
    \frac{1}{t^\alpha}\int_{0^+}^{\frac{\tau'}{t}} du \frac{1}{(1-u)^{\alpha}u}.
\end{align}
For the remaining integral in \eqref{eq: last barB}, the integration interval is compact. Moreover, for any $t>\tau'$ we have $\tau'/t<1$; therefore, the function $(1-u)^{-\alpha}$ remains bounded in the integration domain. Hence, the integral is finite for each $t>\tau'$. In addition, since the remaining integral is multiplied by $t^{-\alpha}$, this term vanishes as $t\to\infty$. Recall that $\lambda$ and $\tau'$ are specific constants associated with each $x\in\mathbb{R}^n$. For $t>\tau'$, every term in $\bar{B}^\alpha(t)$ is continuous and vanishes as $t\to \infty$ for $0<\alpha<1$. Consequently, $\bar{B}^\alpha(t)$ is uniformly bounded.

These upper bounds in \eqref{eq: definition of Balpha(t)}, \eqref{eq: definition graveB}, \eqref{eq: barB definition}, taken together, ensure the following
\begin{equation}
    \frac{|\bar{g}^\alpha_{\mu\nu}|}{M}<B^\alpha(t)<\grave{B}^\alpha(t)\leq \bar{B}^\alpha(t)
\end{equation}
Since $\bar{B}^\alpha(t)$ is uniformly bounded and $M$ is a positive constant, $\bar{g}^\alpha_{\mu\nu}$ is uniformly bounded for $0<\alpha<1$.
Under \emph{Assumption A},

\begin{equation}
    \lim_{t\to\infty} 
    t^{\alpha-1}\bar{g}^\alpha_{\mu\nu}(t,x)=0,
\end{equation}
because $0<\alpha<1$ implies $t^{\alpha-1}\to 0$ and $\bar{g}^\alpha_{\mu\nu}$ is uniformly bounded. Thus, \eqref{eq:equationrewrite} reduces to
\begin{equation}
    \Delta\bar{h}_{\mu\nu}(t,x)=0
    \quad\text{for all sufficiently large } t.
    \label{eq:Laplacian}
\end{equation}
If spacetime is assumed to be asymptotically flat, the solutions of~\eqref{eq:Laplacian} obey
\begin{equation}
\lim_{|x|\to\infty}\bar{h}_{\mu\nu}(t,x)=0.
\end{equation}
It follows from the standard maximum principle that
\begin{equation}
\bar{h}_{\mu\nu}(t,x)=0
\quad\text{for $t\to\infty$}.
\label{eq:finalzero}
\end{equation}
Equivalently, the fractional time operator does not produce a permanent offset in the metric: the system dynamically forgets the disturbance. Mathematically, all memory terms are suppressed by a prefactor $t^{\alpha-1}$ that removes any remaining displacement. The result \eqref{eq:finalzero} demonstrates that

\begin{quote}
\textit{With fractional time operators of the form above, a gravitational waveform generated by a finite-duration burst must decay to zero at late times.}
\end{quote}
Therefore, such models are not sufficient to produce a permanent offset without any flux-balance structures. In some sense, our initial guess of bartering the nonlinearity of Einstein's gravity for the non-locality of the fractional derivative linear theory with intrinsic hereditary features does not yield a proper model of a permanent offset.

In GR, memory arises through global constraints at null infinity. The Christodoulou memory results from integrating the Bondi news:
\begin{equation}
\Delta h_{AB}
\sim
\int_{-\infty}^{+\infty} N_{AB}N^{AB}\,du,
\end{equation}
and is fundamentally tied to the wave's energy flux. In contrast, the fractional model above lacks any analogous symmetry structure or flux law; the decay term forces spacetime to relax back to the trivial flat state. Thus, while fractional damping provides an interesting analytic structure, it cannot serve as a model for true GR memory unless the operator is modified to include nonlinear self-coupling, an asymptotic Bondi structure, or BMS charge flux.

\section{Conclusions and Further Work}

In this work, we investigated whether fractional calculus, with its characteristic long-tail and nonlocal features, can serve as a framework for modeling permanent offset in the field. We examined two toy constructions: ($i$) a sequential Caputo fractional modification of the linearized Einstein field equations and ($ii$) a fractionalized quadrupole formula in which the same operator acts upon the source moment. In both cases, the resulting waveforms displayed history-dependent, memory-like offsets. However, in every numerical experiment, the signal decayed to zero at late times and therefore failed to reproduce the permanent displacement that GR robustly predicts. To verify the late-time decay of the waveforms, we analyze the fractional wave equation in general $n$-dimensional spacetimes, and we have shown that:
\begin{enumerate}
\item The fractional wave equation~\eqref{eq: Fractional 1+n EFE w simplied notation} suppresses late-time fields.
\item Under boundedness and localization assumptions, $\partial_t^\alpha$ terms vanish as $t\to\infty$.
\item The system reduces to the Laplace equation.
\item With asymptotic flatness, $\bar{h}_{\mu\nu}$ must vanish identically.
\end{enumerate}
 Rather than accumulating a permanent displacement, the system erases it. Any memory-compatible fractional modification must violate at least one of the assumptions above.

This outcome demonstrates a key negative result: naive fractionalization of differential operators, while it can mimic hereditary response functions, is not sufficient to capture the permanent offset in the field without any isolation of energy of GWs to null infinity. Importantly, such a null result is not merely technical—it provides a valuable boundary condition on the search for generalized models of gravity. It shows that only models that preserve the flux–balance structure of the BMS symmetries, or an equivalent conservation law tied to the radiated energy flux, can correctly encode permanent gravitational-wave memory. Thus, our study can be viewed as a no-go result for simple fractional generalizations.

The implications are twofold. First, if future theoretical work identifies a fractional framework that preserves flux balance and yields permanent memory, this would represent a genuinely new mathematical structure for radiation in strong gravity. Second, if all such fractional generalizations fail, this reinforces the uniqueness of GR’s prediction of nonlinear memory and highlights the deep role of asymptotic symmetries in the governing radiative phenomena.

Looking ahead, there is strong observational motivation to refine theoretical models of memory. Forecasts suggest that detecting the nonlinear memory may already be possible with extensive event catalogs in the advanced LIGO/Virgo era (e.g., \cite{Huebner2019,Boersma2020}). However, it is more likely to be achieved with next-generation detectors such as the Einstein Telescope and Cosmic Explorer, or with the space-based LISA mission \cite{Favata2009,Favata2010,Ghosh2023}. A confirmed detection would constitute the first direct observational proof of the nonlinear structure of Einstein’s equations in the radiative sector, providing a novel window into strong-field, nonlinear gravity. Moreover, recent work shows that including nonlinear memory in waveform models can even improve parameter estimation by breaking degeneracies (e.g. \cite{Xu2024}).

Several directions emerge from this work:  
\begin{enumerate}

\item Developing hybrid models in which the hereditary flux integral of GR is replaced or augmented by fractional kernels.  
\item Investigating whether fractional calculus can serve as an effective description of modified gravity, including massive gravity theories \cite{KilicarslanTekin2019} that incorporate intrinsic nonlocality.  
\item Exploring the contrast between standard GR, which predicts no memory in higher even-dimensional spacetimes $D>4$ \cite{Garfinkle2017}, and fractional frameworks, which may alter this conclusion.  
\item Establishing the explicit criteria—dimensional consistency, causal structure--that any fractional generalization must meet to be physically viable.  
\end{enumerate}

In summary, our findings underscore that fractional operators alone cannot reproduce gravitational-wave memory; however, the attempt clarifies which structures are required. As the era of precision gravitational-wave astronomy unfolds, both positive and negative results of this kind will be crucial for guiding how we test and extend our understanding of gravity in its most nonlinear regime. One can view our results in the context of infrared finite scattering theory, studied in great detail in \cite{Prabhu2022zcr}, as follows: theories with massless fields defined by the usual wave equations exhibit infrared divergences, which are manifestations of the memory effect. A proper scattering theory for fractional versions of wave equations has not yet been formulated, to the best of our knowledge. However, the absence of memory observed in this work suggests that these theories may not exhibit infrared divergences. This topic needs further investigation.

\clearpage
\appendix
\appendixpage            
\addappheadtotoc         

\numberwithin{equation}{section}
\numberwithin{figure}{section}
\numberwithin{table}{section}
\section
{Nonlinear (Christodoulou) Gravitational-Wave Memory}

Here, to make (\ref{eq: the nonlinear memory}) more transparent, we provide a brief account of the nonlinear gravitational wave memory. We work in asymptotically flat spacetimes with the mostly-plus signature $(-,+,+,+)$ and use retarded Bondi time $u=t-r$ in the wave zone. Bold symbols denote spatial vectors, $\boldsymbol{N}$ is the line-of-sight unit vector to the detector, and ``TT'' means the transverse–traceless projection with respect to~$\boldsymbol{N}$. In units with $G=c=1$, the strain memory ($h$) is dimensionless. Memory is a non-oscillatory, permanent change in the detector-frame strain that remains after a burst or chirp of radiation has passed. If two freely falling test masses are separated by $x^j$ at late times, then their separation contains a step-like offset.  In the TT gauge, the connection between the strain memory and the displacement of the detector masses is readily apparent.
In the TT gauge, the spatial line element between two nearby freely falling test masses with coordinate separation $x^i$ is

\begin{equation}
ds^2=\big(\delta_{ij}+h^{TT}_{ij}\big)\,dx^i dx^j .
\end{equation}
Let $L_0=\sqrt{\delta_{ij}x^i x^j}$ be the unperturbed proper separation and $L$ the perturbed one. To first order in $h^{TT}_{ij}$,

\begin{align}
L
&= \sqrt{(\delta_{ij}+h^{TT}_{ij})\,x^i x^j}
 \simeq L_0\left(1+\tfrac12\,\frac{h^{TT}_{ij}x^i x^j}{L_0^2}\right).
\end{align}
Therefore, the fractional change in proper length is $
\frac{\Delta L}{L_0}=\tfrac12\,h^{TT}_{ij}\,\hat x^i \hat x^j$ with  $\hat x^i:=\frac{x^i}{L_0}$. Identifying the late-time displacement along the original separation direction,
$
\Delta x^i=\frac{\Delta L}{L_0}\,x^i
=\tfrac12\,h^{TT}_{jk}\,\hat x^j \hat x^k\,x^i,
$
and promoting $h^{TT}\to \Delta h^{TT}$ for the net (memory) change, we obtain the standard linearized relation 
\begin{equation}
\Delta x^i \;=\; \tfrac12\, \Delta h^{TT}_{ij}\, x^j,
\qquad
\text{with}\quad
\Delta h^{TT}_{ij} \equiv \lim_{u\to+\infty} h^{TT}_{ij}(u)-\lim_{u\to-\infty} h^{TT}_{ij}(u).
\end{equation}
Equivalently, in geodesic deviation, the \emph{time integral} of the tidal field is nonzero,
\begin{equation}
\int_{-\infty}^{+\infty}\! du\; R_{0i0j}(u)\;\neq\;0,
\end{equation}
so that the passage of the wave leaves a finite late-time displacement rather than a purely oscillatory disturbance. Memory is sourced whenever there is a \emph{net flux of stress–energy to future null infinity} $\mathscr I^+$. There are two contributions to the memory that were discovered at different times. The first one is the linear memory: produced by non-gravitational radiation or ejecta carrying energy–momentum to infinity (e.g., unbound masses, neutrinos), effectively sourced by $T_{uu}^{\mathrm{matter}}$. The second one is the nonlinear (Christodoulou) memory: produced by the energy of the \emph{gravitational waves themselves}, via the GW stress–energy tensor $t^{\mathrm{gw}}_{uu}$. Even if no matter escapes, the burst’s own gravitational energy flux induces a permanent change in the asymptotic shear. In both cases, the cumulative null energy flux fixes a change in the asymptotic Bondi data, which in the detector language is the DC step $\Delta h^{TT}_{ij}$.

As it stands, the memory aspect of the strain may not warrant special attention. But that is a red herring: 
 Memory is a $\omega\!\to\!0$ phenomenon: it depends on the \emph{integrated} energy flux, not on the detailed phasing of the fast oscillations. This makes it robust and largely model-independent once the total radiated energy and its angular pattern are known. The amplitude encodes Bondi mass loss and is tied to BMS supertranslation charges / soft-graviton theorems. Observing it tests deep, symmetry-level predictions of GR rather than only the high-frequency waveform. In gravitational-wave data, memory manifests, not as a step-like offset superposed on the chirp/ringdown. The TT projection tightly constrains its polarization and sky pattern. This offers clean null tests for beyond-tensor polarizations. Because it is DC, memory can be statistically enhanced by stacking many events with aligned signs, providing a complementary science channel even when individual events are marginal. With these points established, we can now write the explicit far-zone expression for the \emph{nonlinear} (Christodoulou) memory.

In the radiation zone, the permanent (DC) change in the observed strain due to \emph{nonlinear} GW memory is (\ref{eq: the nonlinear memory})
, where $dE^{\mathrm{gw}}/(du\,d\Omega')$ is the GW energy flux per unit solid angle in the direction of $\boldsymbol{n}'$, and $r$ is the (luminosity) distance to the source \cite{WisemanWill1991,Thorne1992}.  The TT projector extracts the physical, trace-free tensor transverse to $\boldsymbol{N}$ that an interferometer measures.
Equation~(\ref{eq: the nonlinear memory}) is the retarded far-zone solution of the linearized Einstein equation with an \emph{effective} source given by the GW stress–energy (Isaacson tensor),
\begin{equation}
\Box \bar h_{\mu\nu} \;=\; -16\pi\, t^{\mathrm{gw}}_{\mu\nu}, 
\qquad
t^{\mathrm{gw}}_{uu}(u,\Omega') \;\propto\; \frac{dE^{\mathrm{gw}}}{du\,d\Omega'}.
\end{equation}
Thus, the gravitational field produced by the outgoing radiation \emph{itself} causes a permanent offset in the TT strain once the burst has passed. The inner angular integral folds contributions from all null directions on the past light cone of the detector; the outer time integral via the retarded Green’s function converts a flux into a net change. The physics of the formula  (\ref{eq: the nonlinear memory}) is rich. Let us summarize: When null energy escapes to future null infinity $\mathscr I^+$, the asymptotic (Coulombic) field changes, leaving a lasting shear—the \emph{memory}. This depends only on the emitted energy distribution, not on the fast wave phase.
Memory is a $\omega\!\to\!0$ effect: it is controlled by the integrated flux $\int du\, dE^{\mathrm{gw}}/(du\,d\Omega')$.
The kernel $n'_i n'_j/[1-\boldsymbol{n}'\!\cdot\!\boldsymbol{N}]$ encodes the geometry of the null-cone; emission closer to the line of sight is weighted more strongly. The final TT-projected pattern is quadrupolar.  Restoring $G,c$,
\begin{equation}
\Delta h^{TT}\ \sim\ \frac{4G}{c^{4}}\frac{E^{\mathrm{gw}}}{r}\times\mathcal{O}(1).
\end{equation}
For stellar-mass binaries at hundreds of Mpc radiating a few percent of $Mc^2$, this yields $\Delta h \sim 10^{-23}$--$10^{-22}$.

\section{Memory Effect and Iterative Scheme of Time-Fractional Wave Equation}
Here, we provide a pedagogical derivation of the finite-difference scheme used for the time-fractional wave equation in \eqref{eq: 1+1 Fractional Wave Equation}. The same derivation applies to the fractional quadrupole moment in \eqref{eq: Quadrupole Moment Fractional}. The purpose is to show each step in a way that highlights the physical meaning of the discretization.
We begin with \eqref{eq: 1+1 Fractional Wave Equation}
\begin{equation}
    \frac{\Gamma(2-\alpha)}{t^{1-\alpha}}\partial^\alpha_t\left( \frac{\Gamma(2-\alpha)}{t^{1-\alpha}}\partial^\alpha_tu(t,r) \right)-\partial_r^2u(t,r)=s(t,r).
\end{equation}
with Dirichlet boundaries as $u(t,10)=u(t,30)=0$ for $t>0$ and initial conditions as $u(0^+,r)=0$, $u_t(0^+,r)=0$ on the interval $r\in[10,30]$. To simplify the notation for the governing relation (\ref{eq: 1+1 Fractional Wave Equation}), define the additional function $S(t)$ as
\begin{equation}\label{eq: additional function S(t)}
    S(t) := \frac{\Gamma(2-\alpha)}{t^{1-\alpha}},
\end{equation}
so that the relation (\ref{eq: 1+1 Fractional Wave Equation}) becomes
\begin{equation}\label{eq: Appendix with simple notation 1+1 Fractional Wave}
    S(t)\,\partial^\alpha_t\left[S(t)\,\partial^\alpha_tu(t,r)\right]-\partial^2_r u(t,r)=s(t,r).
\end{equation}
We divide the time interval $[0,T]$ into subintervals $N_t$ of equal size $k=T/N_t$ with nodes $t_n=nk$, $n=0,1,\dots,N_t$. Similarly, we divide the spatial interval $[10,10+L]$ into subintervals $N_r$ of size $h=L/N_r$ with nodes $r_i=10+ih$, $i=0,1,\dots,N_r$. Since we set ($c=1$), the size of the interval must satisfy $L>T$ to avoid boundary reflections. The mesh sizes must be selected so that the errors of \eqref{eq: L1 scheme} and \eqref{eq: central difference approximation for spatial part} are sufficiently small, and the parameters $s_t$, $s_r$, and $w$ of the source \eqref{eq: Source for 1+1 fractional wave equation} have sufficient resolution to obtain accurate results.
To avoid the singularity of $S(t) \propto t^{\alpha-1}$ at $t=0$, we start the time marching at the first positive grid point $t_1=k$ (i.e., $n=1$) and interpret all Caputo derivatives with a lower terminal $0^{+}$.  In particular, $S(t_n)$ is evaluated only for $n \ge 1$, and the $n=1$ history sums (e.g., $K_i^{\,1}$, $H_i^{\,1}$) vanish by definition.
\noindent Following \cite{Murio2008implicit}, we employ the first-order discretization for the Caputo derivative as
\begin{equation}\label{eq: L1 scheme}
    D^\alpha_t f(t_n) \approx \sigma\sum^n_{j=1} w_j (f^{n-j+1}-f^{n-j}),
\end{equation}
where the error is $O(k)$ with the weights and the coefficients
\begin{equation}
    w_j =j^{1-\alpha}-(j-1)^{1-\alpha}, \quad \sigma_{\alpha,k}=\frac{1}{\Gamma(1-\alpha)(1-\alpha)k^\alpha}.
\end{equation}
This formula explicitly shows the “memory” effect: the derivative at step $n$ depends on \emph{all previous time steps}, with power-law weights $w_j$ that implement the fractional kernel. The schematic description of the weights $w_j$ for several $\alpha$ values is given in \cite{Kavvas2022}. To simplify the notation, we denote $\sigma\equiv\sigma_{\alpha,k}$, since $\alpha$ and $k$ values are predetermined before the simulation. 

\noindent We use  \eqref{eq: L1 scheme} to evaluate partial fractional time derivatives on \eqref{eq: Appendix with simple notation 1+1 Fractional Wave} as follows
\begin{equation}
    \partial^\alpha_t u(t_n,r_i) \approx \sigma \sum^n_{j=1}w_j(u^{n-j+1}_i-u^{n-j}_i),
\end{equation}
where the upper indices represent time grid points and the lower indices represent space grid points. To discretize relation \eqref{eq: Appendix with simple notation 1+1 Fractional Wave} at time $t_n$, we define every summation compactly and start from the interior part. Thus, let $q^n_i:=\partial^\alpha_t u(t_n,r_i)$, then use the first-order approximation \eqref{eq: L1 scheme}
\begin{equation}\label{eq: q^n_i}
    q^n_i=\sigma[(u^n_i-u^{n-1}_i)+\sum^n_{j=2}w_j(u^{n-j+1}_i-u^{n-j}_i)].
\end{equation}
Observe that $j=1$ corresponds to a higher term for $u$ in \eqref{eq: q^n_i}. Define $K^n_i$ for the summation in \eqref{eq: q^n_i} as
\begin{equation}\label{eq: Definition of K^n_i}
    K^n_i:=\sum^n_{j=2}w_j(u^{n-j+1}_i-u^{n-j}_i).
\end{equation}
Then, \eqref{eq: q^n_i} becomes
\begin{equation}
    q^n_i=\sigma[(u^n_i-u^{n-1}_i)+K^n_i].
\end{equation}
We define $g^n_i$ for the interior part of the time derivative part in \eqref{eq: Appendix with simple notation 1+1 Fractional Wave} for the discrete case as follows
\begin{equation}
    g^n_i:=S(t_n)\partial^\alpha_t u(t_n,r_i) \quad \Rightarrow \quad g^n_i=S^n q^n_i,
\end{equation}
where $q^n_i$ was defined in \eqref{eq: q^n_i}. Use the first-order approximation in \eqref{eq: L1 scheme} to discretize $\partial^\alpha_t g(t_n,r_i)$,
\begin{equation}\label{eq: g^n_i}
    \partial^\alpha_t g(t_n,r_i) \approx \sigma[(g^n_i-g^{n-1}_i)+\sum^n_{j=2}w_j(g^{n-j+1}_i-g^{n-j}_i)].
\end{equation}
Similarly to $K^n_i$, we define $H^n_i$ for the summation on (\ref{eq: g^n_i}) as \begin{equation}
    H^n_i:=\sum^n_{j=2} w_j(g^{n-j+1}_i-g^{n-j}_i).
\end{equation}
Then, \eqref{eq: g^n_i} becomes
\begin{equation}\label{eq: partial g^n_i}
    \partial^\alpha_t g(t_n,r_i) \approx \sigma[(g^n_i-g^{n-1}_i)+H^n_i].
\end{equation}
We can discretize the interior part of \eqref{eq: Appendix with simple notation 1+1 Fractional Wave} in a simple notational manner as
\begin{equation}\label{eq: first expansion}
    S^n (\partial^\alpha_t g^n_i)-\partial^2_r u^n_i =s^n_i.
\end{equation}
The partial fractional derivative $\partial^\alpha_t g^n_i$ must be written by \eqref{eq: partial g^n_i} and \eqref{eq: first expansion} becomes
\begin{equation}\label{eq: second expansion}
    S^n \sigma[(g^n_i-g^{n-1}_i)+H^n_i]-\partial^2_r u^n_i=s^n_i.
\end{equation}
We insert the definition of $g^n_i$ in  \eqref{eq: g^n_i} to \eqref{eq: second expansion} and expand
\begin{equation}\label{eq: third expansion}
\sigma(S^n)^2 q^n_i-\sigma S^n S^{n-1} q^{n-1}_i+\sigma S^n H^n_i-\partial^2_r u^n_i=s^n_i.
\end{equation}
Similarly, we insert the definition of $q^n_i$ in \eqref{eq: q^n_i} to \eqref{eq: third expansion} as
\begin{equation}\label{eq: bracket expansion one}
    \sigma(S^n)^2\sigma[(u^n_i-u^{n-1}_i)+K^n_i]-\sigma S^n S^{n-1}\sigma[(u^{n-1}_i-u^{n-2}_i)+K^{n-1}_i]+\sigma S^n H^n_i-\partial^2_r u^n_i =s^n_i.
\end{equation}
Expand the relation \eqref{eq: bracket expansion one}
\begin{align}\label{Final expansion}
    s^n_i = & \sigma^2(S^n)^2 u^n_i-\sigma^2 (S^n)^2 u^{n-1}_i+\sigma^2 (S^n)^2 K^n_i-\sigma^2 S^n S^{n-1}u^{n-1}_i \nonumber \\
    &+\sigma^2 S^n S^{n-1}u^{n-2}_i-\sigma^2 S^n S^{n-1} K^{n-1}_i-\sigma S^nH^n_i-\partial^2_ru^n_i.
\end{align}
For the approximation of the spatial derivative, the well-known central difference approximation for the second-order partial derivative is used:
\begin{equation}\label{eq: central difference approximation for spatial part}
    \partial^2_ru(t,r) \approx \frac{u(t,r+h)-2u(t,r)+u(t,r-h)}{h^2}.
\end{equation}
where the error is $O(h^2)$. Then, \eqref{Final expansion} becomes
\begin{align}\label{Final equation}
    s^n_i =& \sigma^2(S^n)^2 u^n_i-\sigma^2 (S^n)^2 u^{n-1}_i+\sigma^2 (S^n)^2 K^n_i-\sigma^2 S^n S^{n-1}u^{n-1}_i \nonumber \\
    &+\sigma^2 S^n S^{n-1}u^{n-2}_i-\sigma^2 S^n S^{n-1} K^{n-1}_i-\sigma S^nH^n_i - \frac{u^n_{i+1}-2u^n_i+u^n_{i-1}}{h^2}.
\end{align}
To solve \eqref{Final equation}, we need to rearrange the equation so that the unknown variables are collected on the left-hand side (LHS) and the known variables are collected on the right-hand side (RHS). After the rearrangement, we obtain
\begin{align}
    \sigma^2(S^n)^2 u^n_i - \frac{u^n_{i+1}-2u^n_i+u^n_{i-1}}{h^2} =& s^n_i+\sigma^2 (S^n)^2 u^{n-1}_i-\sigma^2 (S^n)^2 K^n_i+\sigma^2 S^n S^{n-1}u^{n-1} \nonumber \\
    &-\sigma^2 S^n S^{n-1}u^{n-2}_i+\sigma^2 S^n S^{n-1} K^{n-1}_i+\sigma S^nH^n_i.
\end{align}
The definitions of $K^n_i$ and $H^n_i$ contain only the values of $u^a_b$ where $a<n$. We define the parameter $C^n=\sigma^2(S^n)^2$, and rearrange the LHS to obtain a linear equation. 

\begin{align}
    -\frac{u^n_{i+1}}{h^2}+(C^n+\frac{2}{h^2}) u^n_i - \frac{u^n_{i-1}}{h^2} =& s^n_i+C^n (u^{n-1}_i- K^n_i)+\sigma^2 S^n S^{n-1}u^{n-1}_i-\sigma^2 S^n S^{n-1}u^{n-2}_i \nonumber \\
    &+\sigma^2 S^n S^{n-1} K^{n-1}_i+\sigma S^nH^n_i.
\end{align}
Define everything in the RHS as $RHS^n_i$
\begin{equation}
RHS^n_i := s^n_i + C^n (u^{n-1}_i - K^n_i) + \sigma^2 S^n S^{\,n-1} u^{n-1}_i- \sigma^2 S^n S^{\,n-1} u^{n-2}_i + \sigma^2 S^n S^{\,n-1} K^{\,n-1}_i
+ \sigma S^n H^n_i
\end{equation}
By combining the time-fractional approximation with the finite spatial difference, we obtain a linear system of the form $A u^n = b^n$, where $A$ is a tridiagonal matrix encoding the spatial coupling, and $b^n$ collects known terms from earlier time steps (the memory terms). At each time step, this linear system is solved to update the field $u^n$.

\[
\begin{bmatrix}
C^n+\frac{2}{h^2} & -\frac{1}{h^2} & 0 & \cdots & 0 \\
-\frac{1}{h^2} & C^n+\frac{2}{h^2} & -\frac{1}{h^2} & \cdots & 0 \\
0 & -\frac{1}{h^2} & C^n+\frac{2}{h^2} & \cdots & 0 \\
\vdots & \vdots & \vdots & \ddots & \vdots \\
0 & \cdots & 0 &  -\frac{1}{h^2} & C^n+\frac{2}{h^2}
\end{bmatrix}
\begin{bmatrix}
u^n_2 \\
u^n_3 \\
\vdots \\
u^n_{N_r-1}
\end{bmatrix}
=
\begin{bmatrix}
RHS^n_2 \\
RHS^n_3 \\
\vdots \\
RHS^n_{N_r-1}
\end{bmatrix}
\]
Physically, the iterative scheme demonstrates how the fractional operator induces long-term memory in the evolution: each new step is influenced by the entire history of the source and the field. This is consistent with the hereditary nature of the gravitational memory effect.

The behavior of $u(t,r)$ as $t \to \infty$ can be understood by analyzing the late-time behavior of the iterative scheme. At each time level $t_n$, the update solves $A u^n=b^n$, where $A$ is the tridiagonal matrix with diagonal entries $C^n+2/h^2$, and
\begin{equation}
    C^n=\sigma^2(\Gamma(2-\alpha)t_n^{(\alpha-1)})^2
\end{equation}
where $0<\alpha<1$ is constant. As $n\to \infty$, $t_n^{(\alpha-1)} \to 0$ and hence $C^n \to 0$. For any chosen $N_r$, the matrix $A$ is invertible at each time level $t_n$.

The vector $b^n$ consists of the memory terms ($K^{n}_i$ and $H^n_i$) with the prefactors $C^n$, $\sigma S^n S^{n-1}$, and $\sigma S^n$ together with the source $s^n_i$; since the source is localized in time, $s^n_i=0$ for all sufficiently large $n$. If $u^n_i$, $K^{n}_i$, and $H^n_i$ remain bounded for every $n$, then $S^n \to 0 $ implies $b^n \to 0$. Since $A$ is a nonsingular matrix for any $n$, $u^n =(A)^{-1} b^n \to 0$ as $n \to \infty$, which is consistent with the late-time decay of the offset.

\section*{Acknowledgments}
We thank Prof. Ali Ercan for his valuable discussions and insightful comments on this work, which significantly improved the analysis and raised important questions. TUBITAK-2210-A partially supported S.K.'s work.

\newpage
\printbibliography

\end{document}